\documentclass[]{pasj01}
\onecolumn

\begin{document}
\SetRunningHead{Kitayama et al.}{The SZE at five arc-seconds}
\Received{}
\Accepted{}

\title{The Sunyaev-Zel'dovich Effect at Five Arc-seconds: RX~J1347.5--1145 Imaged by ALMA}

\author{Tetsu \textsc{Kitayama}\altaffilmark{1}, Shutaro 
  \textsc{Ueda}\altaffilmark{2}, Shigehisa
  \textsc{Takakuwa}\altaffilmark{3,4}, 
  Takahiro \textsc{Tsutsumi}\altaffilmark{5}, Eiichiro
  \textsc{Komatsu}\altaffilmark{6,7}, Takuya
  \textsc{Akahori}\altaffilmark{3}, Daisuke
  \textsc{Iono}\altaffilmark{8,9}, Takuma
  \textsc{Izumi}\altaffilmark{8,9}, Ryohei
  \textsc{Kawabe}\altaffilmark{8,9,10}, Kotaro
  \textsc{Kohno}\altaffilmark{11,12}, Hiroshi
  \textsc{Matsuo}\altaffilmark{8,9}, Naomi
  \textsc{Ota}\altaffilmark{13}, Yasushi
  \textsc{Suto}\altaffilmark{12,14}, Motozaku
  \textsc{Takizawa}\altaffilmark{15}, and Kohji
  \textsc{Yoshikawa}\altaffilmark{16}} 
\altaffiltext{1}{Department of
  Physics, Toho University, 2-2-1 Miyama, Funabashi, Chiba 274-8510, Japan; kitayama@ph.sci.toho-u.ac.jp}
\altaffiltext{2}{Institute of Space and Astronautical Science (ISAS),
  Japan Aerospace Exploration Agency (JAXA), 3-1-1 Yoshinodai, Chuo, Sagamihara,  Kanagawa 252-5210, Japan}
\altaffiltext{3}{Department of Physics and Astronomy, Graduate School of Science and Engineering,
Kagoshima University, 1-21-35 Korimoto, Kagoshima, Kagoshima 890-0065, Japan}
\altaffiltext{4}{Academia Sinica Institute of Astronomy and Astrophysics, P.O. Box 23-141, Taipei 10617, Taiwan}
\altaffiltext{5}{National Radio Astronomy Observatory, P.O. Box O, Socorro, NM, 87801, USA}
\altaffiltext{6}{Max-Planck-Institut f\"{u}r Astrophysik, Karl-Schwarzschild Str. 1, D-85741 Garching, Germany}
\altaffiltext{7}{Kavli Institute for the Physics and Mathematics of the Universe (Kavli IPMU), The University of Tokyo, 5-1-5 Kashiwanoha, Kashiwa, Chiba 277-8583, Japan}
\altaffiltext{8}{National Astronomical Observatory of Japan, 2-21-1 Osawa,
Mitaka, Tokyo 181-8588, Japan}
\altaffiltext{9}{The Graduate University for Advanced Studies (SOKENDAI), 
2-21-1 Osawa, Mitaka, Tokyo 181-8588, Japan}
\altaffiltext{10}{Department of Astronomy, 
The University of Tokyo, 7-3-1 Hongo, Bunkyo, Tokyo 113-0033, Japan}
\altaffiltext{11}{Institute of Astronomy, The University of Tokyo, 2-21-1 Osawa, Mitaka, Tokyo 181-0015, Japan}
\altaffiltext{12}{Research Center for the Early Universe, School of Science, 
The University of Tokyo, 7-3-1 Hongo, Bunkyo, Tokyo 113-0033, Japan}
\altaffiltext{13}{Department of Physics, Nara Women's University, Kitauoyanishi-machi, Nara, Nara 630-8506, Japan}
\altaffiltext{14}{Department of Physics, The University of Tokyo, 7-3-1 Hongo, Bunkyo, Tokyo 113-0033, Japan}
\altaffiltext{15}{Department of Physics, Yamagata University, 
1-4-12 Kojirakawa-machi, Yamagata, Yamagata 990-8560, Japan}
\altaffiltext{16}{Center for Computational Sciences, University of Tsukuba, 1-1-1 Tennodai, Tsukuba, Ibaraki 305-8577, Japan}

\KeyWords{cosmology: observations 
-- galaxies: clusters: intracluster medium
-- radio continuum:  general
-- X-rays: galaxies: clusters
-- techniques: interferometric}

\maketitle

\begin{abstract}
We present the first image of the thermal Sunyaev-Zel'dovich effect
(SZE) obtained by the Atacama Large Millimeter/submillimeter Array
(ALMA). Combining 7-m and 12-m arrays in Band 3, we create an SZE map
toward a galaxy cluster RX~J1347.5--1145 with 5 arc-second resolution
(corresponding to the physical size of 20$h^{-1}$kpc), the highest
angular and physical spatial resolutions achieved to date for imaging
the SZE, while retaining extended signals out to 40 arc-seconds.  The
1$\sigma$ statistical sensitivity of the image is 0.017 mJy/beam or
0.12 mK$_{\rm CMB}$ at the 5 arc-second full width at half maximum.
The SZE image shows a good agreement with an electron pressure map
reconstructed independently from the X-ray data and offers a new probe
of the small-scale structure of the intracluster medium. Our results
demonstrate that ALMA is a powerful instrument for imaging the SZE in
compact galaxy clusters with unprecedented angular resolution and
sensitivity. As the first report on the detection of the SZE by ALMA,
we present detailed analysis procedures including corrections for the
missing flux, to provide guiding methods for analyzing and
interpreting future SZE images by ALMA.

\end{abstract}

\section{Introduction}

The Sunyaev-Zel'dovich effect (SZE, \cite{Sunyaev72}), inverse Compton
scattering of the cosmic microwave background (CMB) photons off hot
electrons, offers a powerful probe of cosmic plasma up to high
redshifts (see \cite{Rephaeli95, Birkinshaw99, Carlstrom02,
  Kitayama14} for reviews).  The surface brightness of the SZE is
independent of the source redshift $z$ for given electron density
$n_{\rm e}$ and temperature $T_{\rm e}$ and is proportional to $n_{\rm
  e} T_{\rm e}$, whereas that of X-rays varies as $n_{\rm
  e}^2(1+z)^{-4}$ with only weak dependence on $T_{\rm e}$. The SZE
can hence be a unique tool for studying the physics of the
intracluster medium, e.g., by detecting shocks (pressure gaps) and
very hot gas associated with subcluster mergers
\citep{Komatsu01,Kitayama04,Korngut11}.  The advent of large-area
surveys by the South Pole Telescope (SPT) (e.g.,
\cite{Spt09,Spt10,Spt11,Spt13a}), the Atacama Cosmology Telescope
(ACT) (e.g., \cite{Act10,Act11,Act13a}), and the Planck satellite
(e.g., \cite{Planck_earlysz,Planck13a}) has enhanced the sample of
galaxy clusters observed via the SZE by more than an order of
magnitude over the past decade.  The caveats of existing observations
are limited angular resolution ($>1'$ in the above mentioned surveys)
and sensitivity. Single-dish measurements by the MUSTANG bolometer
array have achieved currently the highest angular resolution of $9''$
full width at half maximum (FWHM) for the SZE maps (e.g.,
\cite{Mason10,Korngut11,Romero15,Young15}), while they are still
challenged by point source and atmospheric contamination.
Interferometers offer a complementary tool with good control of
systematic noise and capability of separating compact sources from the
SZE, albeit reduced sensitivity for the sources more extended than the
baseline coverage (e.g., \cite{Jones93, Carlstrom96, AMI06,
  Muchovej07, Wu09}).  A recent SZE map obtained by CARMA has a
synthesized beam with $10.6'' \times 16.9''$ \citep{Plagge13}.

With a combination of 7-m and 12-m arrays in Band 3, the Atacama
Large Millimeter/submillimeter Array (ALMA) serves as the first
instrument to resolve the SZE with an angular resolution of $5''$, as
predicted by detailed imaging simulations \citep{Yamada12} using
hydrodynamic simulation data \citep{Takizawa05,Akahori12}. Among
currently available frequency bands of ALMA, Band 3 is the most
suitable for the SZE imaging owing to the largest field-of-view, the
lowest system temperature, and minimal contamination by synchrotron
and dust emission.  Given that the Total Power Array is still
unavailable for continuum observations by ALMA, viable targets are
limited to compact distant galaxy clusters.

In this paper, we present the first measurement of the SZE by
ALMA. The target is a galaxy cluster RX J1347.5--1145 at $z=0.451$.
Owing to its brightness and compactness, RX J1347.5--1145 is a prime
target for imaging observations by the current configuration of
ALMA. A number of SZE measurements have been made for this cluster in
the past
\citep{Komatsu99,Pointecouteau99,Komatsu01,Pointecouteau01,Reese02,
  Carlstrom02,Kitayama04,Benson04,Zemcov07,Mason10,Korngut11,Zemcov12,
  Plagge13,Adam14,Sayers16}. In particular, the Nobeyama Bolometer
Array (NOBA; \cite{Komatsu01}) detected a prominent substructure that
was not expected from regular morphology of this cluster in the
soft-band X-ray image by ROSAT \citep{Schindler97}.  The presence of
the substructure was confirmed with subsequent X-ray data by Chandra
\citep{Allen02,Johnson12} and XMM-Newton \citep{Gitti04} as well as
SZE maps by MUSTANG \citep{Mason10,Korngut11}, CARMA \citep{Plagge13},
and NIKA \citep{Adam14}. The inferred temperature of the substructure
exceeds 20 keV and is appreciably higher than the mean temperature of
the cluster $\sim 13$ keV \citep{Kitayama04,Ota08}; this accounts for
the fact that the substructure was more obvious in the SZE map than
the X-ray surface brightness image. The disturbed feature was also
observed by the radio synchrotron observations
\citep{Gitti07,Ferrari11} and gravitational lensing maps (e.g.,
\cite{Miranda08,Bradac08,Koehlinger14}). These previous results
indicate that the cluster is undergoing a merger, but its exact nature
such as geometry and dynamics of the collision is still unclear
\citep{Johnson12,Kreisch16}. The ALMA Band 3 observation of RX
J1347.5--1145 is crucial not only for better understanding this
particular galaxy cluster but also for testing the capability of ALMA
in observing the SZE against a range of independent datasets available
for this well-studied system.

This paper is organized as follows. Section \ref{sec-obs} describes
the observations and calibration. Section \ref{sec-image} presents
details of the imaging analysis including point source subtraction and
deconvolution. The results are validated against the X-ray data,
realistic imaging simulations, and previous high-significance SZE
measurements in Section \ref{sec-implication}. Finally, our
conclusions are summarized in Section \ref{sec-conc}.  Throughout the
paper, we adopt a standard set of cosmological density parameters,
$\Omega_{\rm M}=0.3$ and $\Omega_{\rm \Lambda}=0.7$. We use the
dimensionless Hubble constant $h\equiv H_0/(100 \mbox{km/s/Mpc})$;
given controversial results on the value of $h$
(e.g., \cite{Planck15,Riess16}), we do not fix it unless stated otherwise.
In this cosmology, the angular size of 1$''$ corresponds to the
physical size of 4.04 $h^{-1}$kpc at the source redshift
$z=0.451$. The errors are given in 1$\sigma$ and the coordinates are
given in J2000.

\section{Observations and Calibration}
\label{sec-obs}

RX J1347.5--1145 was observed by ALMA in Band 3 (Project code:
2013.1.00246.S).  The data were taken with the 12-m array and the 7-m
array at 13 separate Execution Blocks listed in Table \ref{tab-obs}
between August 2014 and January 2015. Compact configurations were
adopted to cover the projected baseline ranges of $7.4 - 49$~m and
$12.5 - 347$~m for the 7-m array and the 12-m array, respectively.  In
all the Execution Blocks, the dual-polarization Time Division Mode was
adopted with the central observing frequency of 92~GHz and an
effective band width of 7.5GHz; the data were taken at four spectral
windows with widths $\sim 2$ GHz centered at 85, 87, 97, and 99 GHz,
respectively.  The effective primary beam FWHMs at 92~GHz are $62''$
and $107''$ for 12-m array and 7-m array, respectively. At these
frequencies, the above mentioned baseline ranges correspond to the
$uv$ distances of $2.1 - 16.3$ k$\lambda$ and $3.5 - 116$ k$\lambda$
for the 7-m array and the 12-m array, respectively. The parameters of
the observed maps are summarized in Table \ref{tab-param}.

\begin{table}
  \caption{Execution Blocks of Band 3 observations for RX
    J1347.5--1145.  The J1337--1257 flux density at 92GHz was obtained
    by a power-law fit to the flux at 85, 87, 97, 99 GHz in the ALMA
    QA2 report, based on the original flux scale using the flux
    calibrator shown in the table. In EB7-9 and EB12-2, the flux
    scales using Ceres were not used in the subsequent analysis and
    replaced by those of EB7-8 and EB12-3,
    respectively. }\label{tab-obs}
  \begin{center}
    \begin{tabular}{ccccccc}
      \hline
  ID & Array & Date & Number of & On-source time & 
Flux Calibrator  &  J1337--1257 flux density \\
     &      &     &  Antennas  & [min]    & & at 92GHz [Jy]\\ \hline
   EB7-1 & 7-m & 2014-08-16  & 10 & 39.43 & Mars & $4.430 \pm 0.003$\\
   EB7-2 & 7-m & 2014-08-17 & 10 & 39.43 & Mars & $4.745 \pm 0.003$ \\
   EB7-3 & 7-m & 2014-08-17 & 10 & 19.97 & Mars & $4.582 \pm 0.002$\\
   EB7-4 & 7-m & 2014-08-17  & 9 & 39.43 & Mars & $4.290 \pm 0.003$\\
   EB7-5 & 7-m & 2014-12-06 & 7 & 39.43 &  J1337--1257& $5.041 \pm 0.003$\\
   EB7-6 & 7-m & 2014-12-11 & 9 & 39.43 & Callisto & $4.416 \pm 0.031$\\
   EB7-7 & 7-m & 2014-12-15 & 8 & 39.43 & J1337--1257 & $4.808 \pm 0.001$\\
   EB7-8 & 7-m & 2014-12-28 & 8 & 39.43 & Callisto & $5.102 \pm 0.067$\\
   EB7-9 & 7-m & 2014-12-28  & 8 & 39.43 & (Ceres) & ($3.940 \pm 0.016$)\\
   EB12-1 & 12-m & 2014-12-15 & 41 & 34.13 & Ganymede & $4.358 \pm 0.002$\\
   EB12-2 & 12-m & 2014-12-29 & 39 & 40.42 & (Ceres) & ($5.621 \pm 0.002$) \\
   EB12-3 & 12-m & 2014-12-30 & 39 & 40.42 & J1337--1257 & $4.649 \pm 0.001$\\
   EB12-4 & 12-m & 2015-01-04 & 40 & 40.42 & J1337--1257 &  $4.916 \pm 0.001$\\
      \hline
    \end{tabular}
  \end{center}
\end{table}

\begin{table} 
  \caption{Parameters of observed maps.}\label{tab-param}
  \begin{center}
    \begin{tabular}{ccc}
      \hline
  Parameters & 12-m array & 7-m array \\ \hline 
  Central frequency & 92~GHz & 92~GHz \\
  Band widths & 7.5~GHz & 7.5~GHz \\
  Primary beam FWHM at the central frequency & $62''$ &  $107''$ \\
  Number of pointings & 7  & 7 \\
  Baseline coverage & $3.5 - 116$ k$\lambda$ & $2.1 - 16.3$ k$\lambda$ \\
  Weighting & natural & natural \\
Synthesized beam FWHMs & $4.1''\times 2.4''$ 
& $20.5'' \times 11.1''$  \\ 
Synthesized beam position angle & $84.1^{\circ}$ &$88.1^{\circ}$  \\ 
Average 1$\sigma$ noise & 0.012 mJy/beam & 0.083 
mJy/beam \\ 
      \hline
    \end{tabular}
  \end{center}
\end{table}

The observing field has a diameter of $\sim 90''$ centered at
(\timeform{13h47m30.54s},\timeform{-11d45m19.40}), which is about
$10''$ south of the X-ray center of RX J1347.5--1145 and is closer to
the peak of the offset SZE signal \citep{Komatsu01}. The field is
covered with 7 hexagonal mosaic pointings with an equal spacing of
$34.2''$ by both arrays. This is approximately the Nyquist spacing for
the 12-m array and achieves much denser sampling for the 7-m array.

Calibration of the raw visibility data taken with the 12-m array
  was performed using the Common Astronomy Software Applications
  (CASA; \cite{McMullin07}) program version 4.3.1 as implemented in a
  standard reduction script for ALMA Cycle 2 data. The data taken with the
  7-m array were calibrated by the ALMA pipeline using the CASA
  version 4.2.2. We use the data produced from the second stage of
ALMA's Quality Assurance process (QA2) for both the 12-m array and the
7-m array.

A quasar J1337--1257 was monitored in all the execution blocks; it is
used as a phase calibrator as well as a bandpass calibrator throughout
this project. Several solar system objects (Mars, Callisto,
Ceres, and Ganymede; see Table \ref{tab-obs}) were chosen by the
observatory as primary flux calibrators, although the data of Ceres
were not used in the end for the reason described below. We used these
calibrators to determine the flux density of J1337--1257, which was
then used to compare the flux scales of different execution blocks.

The Butler-JPL-Horizons-2012 Model (Butler 2012) was used for
determining the absolute flux density of solar system objects. The
absolute flux densities of J1337--1257 in EB7-5, EB7-7, EB12-3, and
EB12-4 were determined from measurements of the solar system objects
by the observatory on 2014-12-07, 2014-12-18, and 2015-01-17,
available in the ALMA Calibrator Source
Catalogue\footnote{https://almascience.nao.ac.jp/sc/}.  Given that the
catalogued measurements were more than 10 days away from EB12-3 and
EB12-4, we additionally checked that the flux scales of EB12-3 and
EB12-4 were consistent with three calibration measurements of
J1337--1257 in Band 3 executed on 2015-01-06, the closest in time
available in the ALMA archive, using Ganymede, Pallas, and J1427--421
as flux calibrators (Project codes: 2013.1.00120.S and
2013.1.01312.S).

Flux equalization was applied during QA2 to some of the execution
blocks close in time.  Observations in August, 2014 (EB7-1, EB7-2,
EB7-3, and EB7-4) were performed within 27 hours and scaled together
using the mean of the absolute flux scales from the four execution
blocks.  For EB7-8 and EB7-9 executed within 6 hours, the absolute
flux scale of the former was adopted, because the flux calibrator of
the latter (Ceres) was weak. EB12-2 and EB12-3 performed within 24
hours were also scaled together adopting the flux scale of the latter,
because the flux calibrator of the former (Ceres) was weak.

The variance of the J1337--1257 flux density listed in Table
\ref{tab-obs} is much larger than the errors quoted in individual
execution blocks, suggesting the presence of underlying systematics of
the absolute flux scale as well as potential time variability of
J1337--1257 itself.
For 11 execution blocks excluding EB7-9 and EB12-2 (for which the flux
scale was replaced with those of other execution blocks as mentioned
above), the standard deviation of the J1337--1257 flux density at
92~GHz relative to the mean is $5.7\%$. If we separate the data taken
in August, 2014, and the others, it is $3.8\%$, and $5.7\%$,
respectively, in each data set.  If we separate the data taken by the
7-m array and the 12-m array, it is $6.0\%$, and $4.9\%$,
respectively.  Finally, in each of the four spectral windows for the
11 execution blocks, it is $5.7,~6.4,~5.6,~5.7\%$ at 85, 87, 97, 99
GHz, respectively. These values correspond to the variance of the flux
scales determined by a range of calibrators used, both solar system
objects and J1337--1257. Given these results, we estimate the flux
calibration error in the present analysis to be $6\%$, apart from any
unknown systematics that may offset the entire calibration
measurements.  This precision is consistent with the expected
performance of Cycle 2 described in the ALMA Technical Handbook
\footnote{https://almascience.nrao.edu/documents-and-tools/cycle4/alma-technical-handbook}.

\section{Imaging Analysis}
\label{sec-image}

We performed imaging analysis using CASA version 4.5.0 on the
visibility data produced by the ALMA QA2 process. Visibility weights
are assigned according to the standard calibration procedure described
in Section \ref{sec-obs} using CASA versions 4.3.1 and 4.2.2 for the
data taken by the 12-m array and the 7-m array, respectively.  Natural
weighting is adopted to maximize sensitivity to an extended signal. We
use the multi-frequency synthesis mode for spectral gridding and the
mosaic mode for imaging. Deconvolution is done using the Multi-Scale
CLEAN algorithm (see Section \ref{sec-deconv} for details).  All the
results presented have been corrected for primary beam attenuation.
Whenever smoothing is done, a Gaussian kernel is used and the referred
size corresponds to FWHM.

\subsection{Dirty Maps}
\label{sec-dirty}

\begin{figure}
 \begin{center}
  \includegraphics[width=16.5cm]{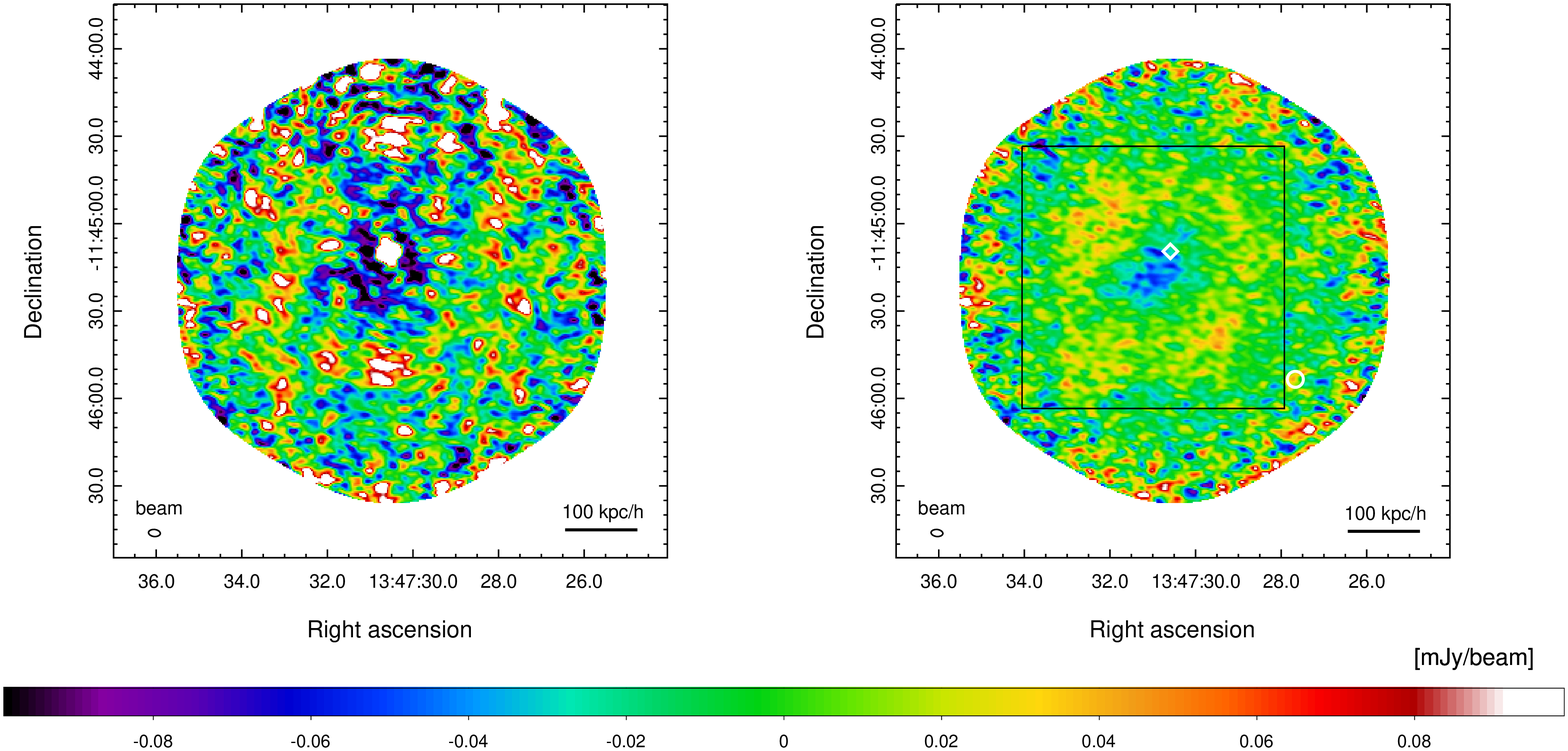} 
  \includegraphics[width=16.5cm]{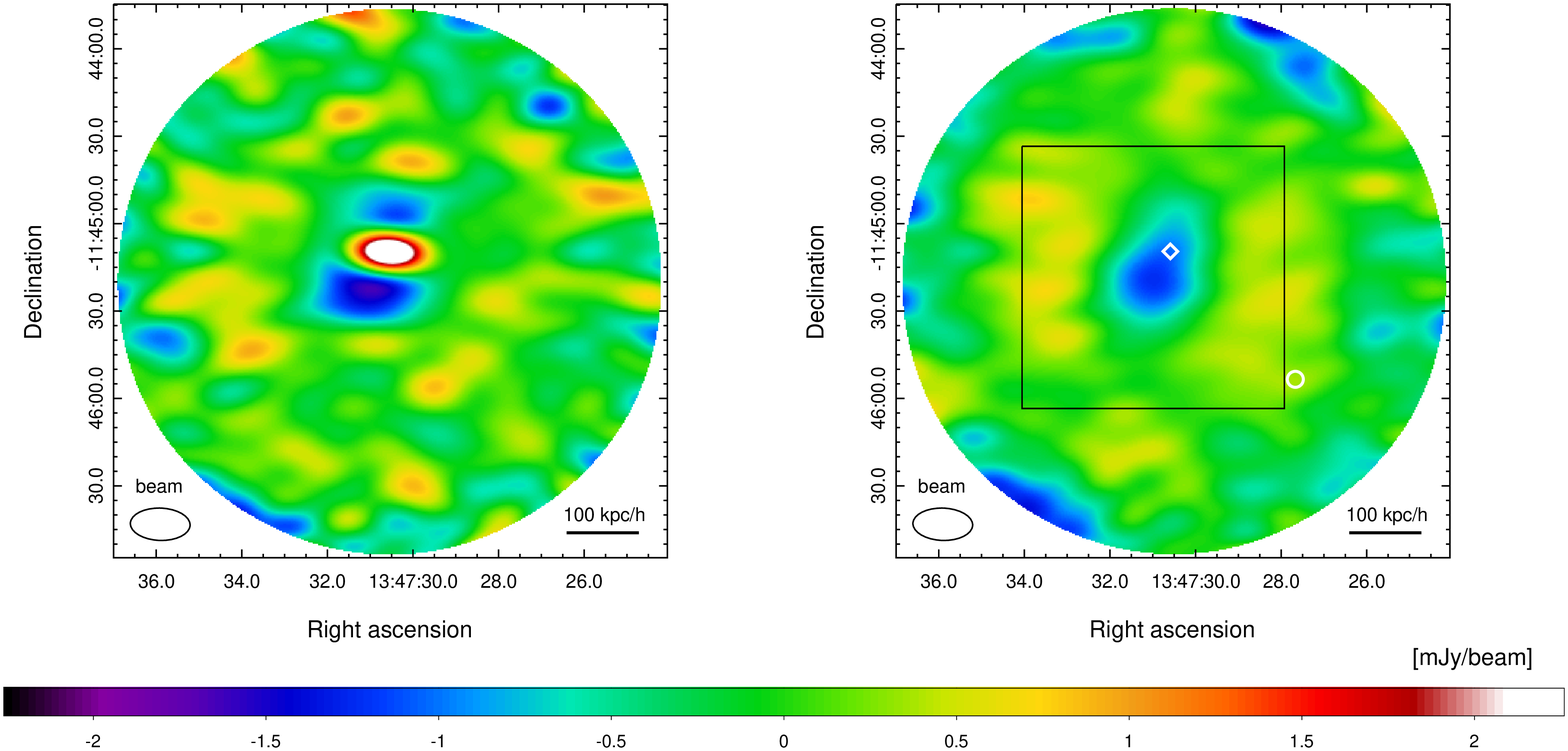} 
 \end{center}
\caption{Dirty maps of the data taken by the 12-m array (top) and the
  7-m array (bottom) before (left) and after (right) the central AGN
  is subtracted. The synthesized beams have $4.1''\times 2.4''$ FWHMs
  with a position angle $84.1^{\circ}$ and $20.5''\times 11.1''$ with
  a position angle $88.1^{\circ}$ in the top and bottom panels, as
  indicated at the bottom left of each panel. In the right panels, the
  positions of the central AGN (subtracted) and another radio source
  detected by VLA and SCUBA (unsubtracted and excluded from the
  analysis) are marked by a diamond and a circle, respectively. The
  square indicates a $90'' \times 90'''$ box centered on the SZE peak
  over which Figures \ref{fig-dirty2} and \ref{fig-sz} are 
  plotted. The ranges of the color scale in all the panels correspond
  to the same brightness in units of Jy/arcsec$^2$. }
\label{fig-dirty1}
\end{figure}

Figure \ref{fig-dirty1} shows dirty maps produced directly from the
calibrated visibility data for each of the 12-m array and the 7-m
array. The synthesized beam parameters are listed in Table
\ref{tab-param}.  The pixel size is $0.5''$. A positive compact source
is present near the field center. This source is a known AGN hosted in
the Brightest Central Galaxy of RX J1347.5--1145 and its flux density
has been reported at a wide range of frequencies in literature (see
Figure 1 of \cite{Sayers16} for a compiled spectrum and references).
The baseline coverage of ALMA allows us to identify the position and
the flux density of this source almost simultaneously with the SZE,
minimizing the impact of source variability and potential systematics
of using external datasets. No other compact sources are detected at
more than the $4\sigma$ statistical significance in the observing
field. We present full details of the source subtraction procedure as
well as the flux limit on another known radio source in the field in
Section \ref{sec-source}.

\begin{figure}
 \begin{center}
  \includegraphics[width=16.5cm]{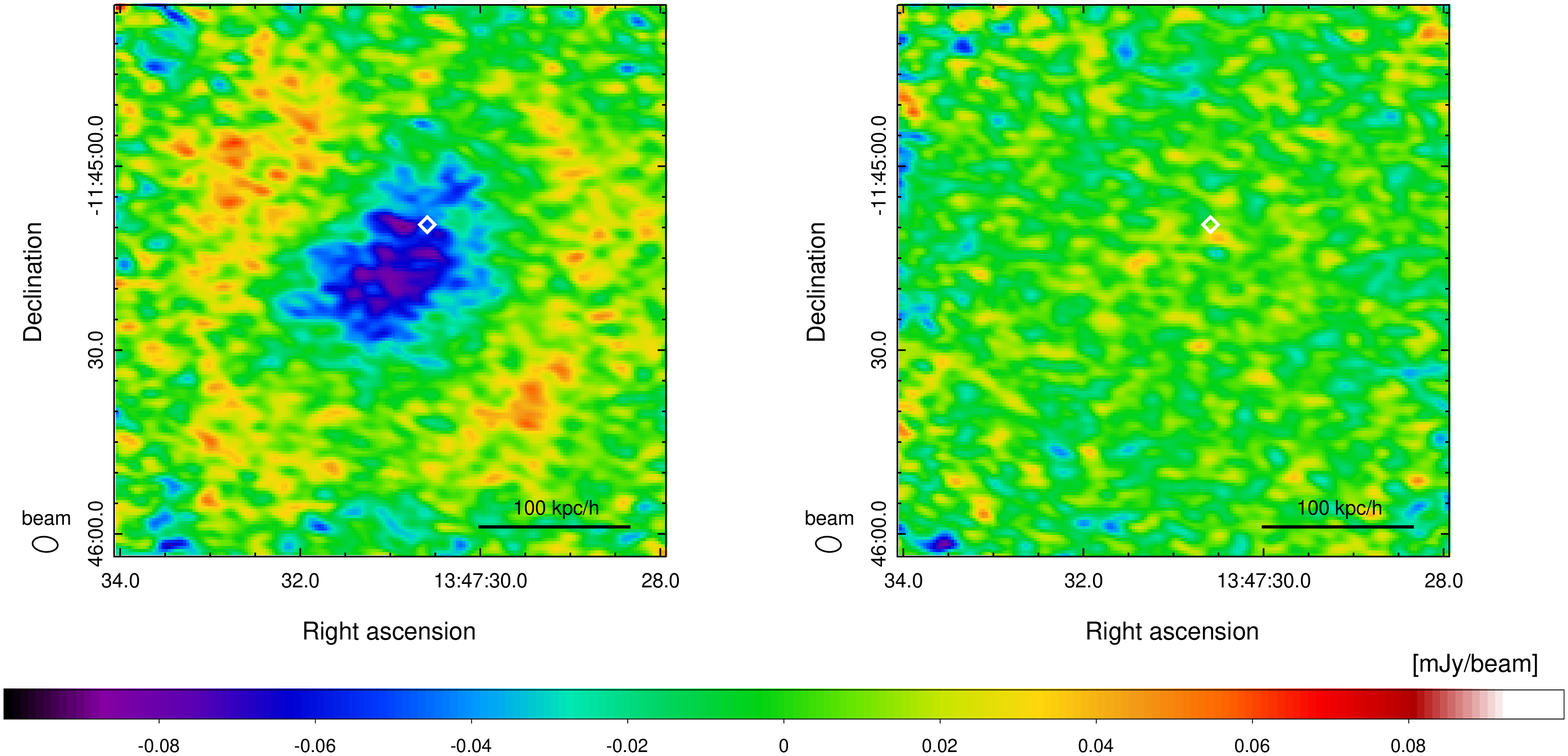} 
 \end{center}
\caption{Dirty map of RX J1347.5--1145 after the central AGN is
  subtracted and the data taken by the 12-m array and the 7-m array
  are combined (left).  For reference, a difference map created by
  dividing the data set in half is also shown (right).  The position
  of the subtracted central AGN is marked by a diamond. The
  synthesized beam has $4.1''\times 2.5''$ FWHMs with a position angle
  $84.1^{\circ}$ as indicated at the bottom left of each panel. The
  ranges of the color scale correspond to the same brightness in units
  of Jy/arcsec$^2$ as in Figure \ref{fig-dirty1}. }
\label{fig-dirty2}
\end{figure}

Once the compact central source is removed, the extended negative
signal of the SZE becomes more apparent.  Figure \ref{fig-dirty2}
further illustrates that the dynamic range of the signal is enhanced
significantly once the data taken by the 12-m array and the 7-m array
are combined.  The synthesized beam of the combined map has
$4.1''\times 2.5''$ FWHMs with a position angle $84.1^{\circ}$.
Positive-valued pixels surrounding the central decrement are due to
side-lobes of the dirty beam, which will be corrected by deconvolution
in Section \ref{sec-deconv}.

To measure the noise levels excluding the extended signal, we also
created a difference map by dividing the data set in half, taking a
difference between their dirty images, and dividing it by 2 to correct
for the reduction of the integration time (right panel of Figure
\ref{fig-dirty2}).  The rms values on the difference map are $0.011,
~0.010, ~0.012, ~0.015$, and $0.023$ mJy/beam within diameters of $30, ~60,
~90, ~120$, and $150''$ around the field center, respectively; the noise
level is nearly constant except at the map edge. We have repeated a
similar analysis on the 12-m array data and the 7-m array data
separately to find that the rms values within the diameter of $90''$
are 0.012 mJy/beam, and 0.083 mJy/beam, respectively.

\subsection{Point Source Subtraction}
\label{sec-source}

We determined the position and the flux density of the central AGN at
our observing bands using long baseline data at the $uv$ distance
larger than $30$k$\lambda$. This cutoff $uv$ distance was chosen to
fully separate the compact source from the extended signal beyond
$\sim 5''$ as well as to retain an ample signal-to-noise ratio. The
source is consistent with a point source if an image was created from
visibilities above this cutoff.

Given simplicity of the source shape, we fitted a point source model
to the $uv$ data from the central pointing (without mosaicing) varying
both the flux density and the position. To examine the spectrum of the
source, we performed the fit separately in 84-88 GHz and 96-100 GHz,
i.e., in upper and lower two spectral windows. Fitting the data in
EB12-1, EB12-2, EB12-3, and EB12-4 simultaneously, we find $F_\nu=4.16
\pm 0.03 \pm 0.25$ mJy and $3.96 \pm 0.03 \pm 0.24$ mJy at the central
frequencies of 86 GHz and 98 GHz, respectively. The first errors are
statistical and the second ones are systematic errors from the flux
calibration.  The best-fit source position is
(\timeform{13h47m30.62s},\timeform{-11d45m09.5}) at both frequencies
with a statistical error of less than $0.02''$. The spectral index of
the source inferred from the two measurements over a narrow range
between 86 GHz and 98 GHz is $-0.38 \pm 0.12$. If the fit is performed
for smaller spectral bins, the flux density changes only within the
statistical errors.  This is also consistent with the spectral index
inferred above; the flux density is expected to change by less than
about $1\%$ over the frequency difference of $2$~GHz in our observing
band.  We hence did not subdivide the frequency bins further.  The
best-fit point source models described above were subtracted from the
visibility data taken by the 12-m array and the 7-m array, separately
in each of the spectral windows covering 84--88 GHz and 96--100 GHz.

When we fit each of EB12-1, EB12-2, EB12-3, and EB12-4 separately, we
find that the best-fit flux of the central AGN varies around the above
mentioned values by $\pm 4.3\%$ and $\pm 6.4\%$ at 86 GHz and 98 GHz,
respectively, whereas the statistical errors are about $\pm 1.5\%$.
The variation is correlated with that of the J1337--1257 flux listed
in Table \ref{tab-obs}; the linear correlation coefficients are 0.993
and 0.999, respectively. This suggests that systematic uncertainty of
the central AGN flux is indeed dominated by the calibration error.

To test consistency of the normalization of the 7-m and 12-m array
data, we fitted a point source model to the visibilities taken by the
7-m array between 7 and 16 k$\lambda$, the longest $uv$ distances
covered by this array.  Given poorer statistics of the 7-m array data,
all the spectral windows between 84 GHz and 100 GHz are fitted
together fixing the source coordinate at the best-fit position from
the 12-m array data. From the data in all the execution blocks for the
7-m array, the fitted value of the AGN flux density is $4.09 \pm 0.17
\pm 0.25$ mJy. If we separate the data taken in 2014 August and 2014
December, they are $4.22 \pm 0.24 \pm 0.25$ mJy and $3.95 \pm 0.24 \pm
0.24$ mJy, respectively. Statistical errors are too large to examine a
correlation with the J1337--1257 flux density for the 7-m array. For
comparison, a similar fit to the 12-m array data between 7 and 16
k$\lambda$ yields $4.04 \pm 0.03 \pm 0.24$ mJy.  The flux values
measured by the two arrays are thus consistent within the statistical
and calibration errors.

In addition to the calibration error of $6\%$ on the total SZE flux,
the intensity of the source-subtracted map is subject to a
position-dependent error from the adopted source flux. At the
subtracted source position, this error is estimated to be at most
$3\%$ of the source flux ($\sim 4$mJy), i.e., $0.12$ mJy/beam, after
the data from 4 and 9 execution blocks are combined for the 12-m array
and the 7-m array, respectively.
The error becomes smaller than the noise level of 0.012 mJy/beam at
$2.9''$ away from the source for an average beam FWHM of $3.2''$ in
the present case.

Note that the flux density of the central AGN measured by ALMA in 2014
December and 2015 January are lower than $4.9\pm 0.1$ mJy measured by
the 23 element array of CARMA in 2011 February at the central
frequency of 86~GHz \citep{Plagge13}; the fitted source position is in
good agreement with that of CARMA. A part of the difference may be
ascribed to a long-term variability of the AGN over nearly four-year
period as well as to calibration between ALMA and CARMA. Even if we
further allow for an additional bias of +1~mJy for the central source
flux, the biased intensity in the source-subtracted map drops below
the noise level at $4.0''$ away from the source.

We hence conclude that the flux calibration error hardly affects {\it
  morphology} of the source-subtracted map except in the vicinity of
the source position. Still, its impact on the total flux should be
treated with care.

There is another known radio source in the field of RX J1347.5--1145
at (\timeform{13h47m27.7s},\timeform{-11d45m53.4}) with the 1.4 GHz
flux density of $4.50 \pm 0.06$ mJy measured by VLA \citep{Gitti07}
and the 350 GHz flux density of $15.4 \pm 5.3$ mJy measured by SCUBA
\citep{Kitayama04}. This source is located at $\sim 60''$ away from
the central AGN (a circle in Figure \ref{fig-dirty1}). Being at the
map edge, this source is not clearly detected in the present ALMA data
while there is a hint of a positive signal at about the $ 3\sigma$
statistical significance. The 92~GHz flux density measured within the
diameter of $10''$ around the reported VLA position is $0.17 \pm 0.06$
mJy. We do not subtract this source and simply exclude its position
from the analysis presented in this paper.

\subsection{Deconvolution of Extended Signal}
\label{sec-deconv}

To correct for incomplete sampling of the visibility data, we
performed image deconvolution using the Multi-Scale CLEAN algorithm
implemented in CASA. This method models an image by a sum of Gaussians
of various sizes and is shown to recover much higher fraction of flux
for extended sources than the conventional CLEAN algorithm
\citep{Cornwell08,Rich08}. It also has an advantage over the Maximum
Entropy Method (e.g., \cite{Narayan86, Sault90}) in that it is free
from the positivity constraint.

There are some free parameters in Multi-Scale CLEAN, such as the sizes
of Gaussian components used for deconvolution and the scale bias
parameter by which residuals at different scales are weighted
\citep{Cornwell08}. Optimal choices of these parameters depend on the
target and the instrument in consideration. Having searched for a
combination of parameters that gives minimal residuals as well as the
maximal recovered flux, we adopt the scales [$0, ~5'', ~8'', ~13'',
  ~20'', ~30'', ~45''$] with the small-scale bias parameter of 0.4 as
our fiducial choice. The scale $0$ corresponds to the size of a
synthesized beam, i.e., the scale used in the conventional
(single-scale) CLEAN algorithm. We also use a loop gain of 0.05
  suitable for diffuse emission and a flux threshold of 0.03 mJy ($2.5
  \sigma$); the flux threshold is reached by less than 10,000
  iterations in all the cases presented in this paper. 

\begin{figure}
 \begin{center}
  \includegraphics[width=9cm]{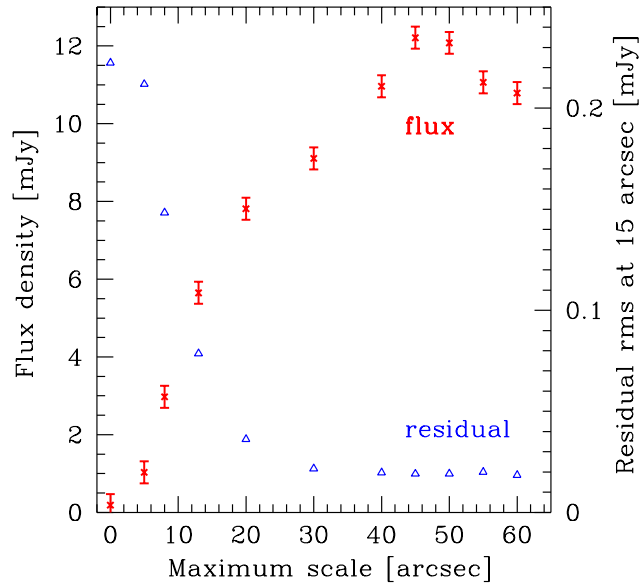} 
 \end{center}
\caption{Recovered flux and residuals of deconvolved images versus the
  maximum scale of Gaussians used in Multi-Scale CLEAN.  Wherever
  smaller than the scale shown in the figure, Gaussians with FWHMs of
  0, $5''$, $8''$, $13''$, $20''$, and $30''$ are also used in the
  deconvolution, i.e., the smaller scale values are fixed for
  comparison. The plotted residual is the rms value measured on a
  residual map after being corrected for primary beam attenuation and
  smoothed with a Gaussian kernel to a fixed resolution of $15''$
  FWHM. Both flux and residuals are measured within a diameter of
  $90''$ around the field center.  }
\label{fig-scale}
\end{figure}

Figure \ref{fig-scale} illustrates how the recovered flux and
residuals change as we successively add a larger scale in the
deconvolution.  To examine the sensitivity to an extended signal, the
residual maps are smoothed by a Gaussian kernel to a resolution of
15$''$ FWHM and the rms values within a diameter of $90''$ around the
field center are plotted. They are compared with the flux within the
same region of the deconvolved image.  Converged results are obtained
by taking the largest scale to be between $40''$ and $50''$,
corresponding to the maximum recoverable scale for the shortest
baseline length of 2.1 k$\lambda$ in the present data.
We have also checked that the deconvolved results are insensitive to
the value of the small-scale bias parameter as long as its value is
less than 0.8.

While the above procedure gives robust reconstruction of the observed
signal up to a spatial scale of $\sim 40''$, flux at larger scale is
still lost in currently available ALMA data with no total power
measurements.  In what follows, we focus on the results that can be
derived solely at $< 40''$ and discuss separately the degree of
missing flux by means of simulations in Section \ref{sec-sim}.

\subsection{The Sunyaev-Zel'dovich Effect Map of RX J1347.5--1145}
\label{sec-sz}

\begin{figure}
 \begin{center}
  \includegraphics[width=9cm]{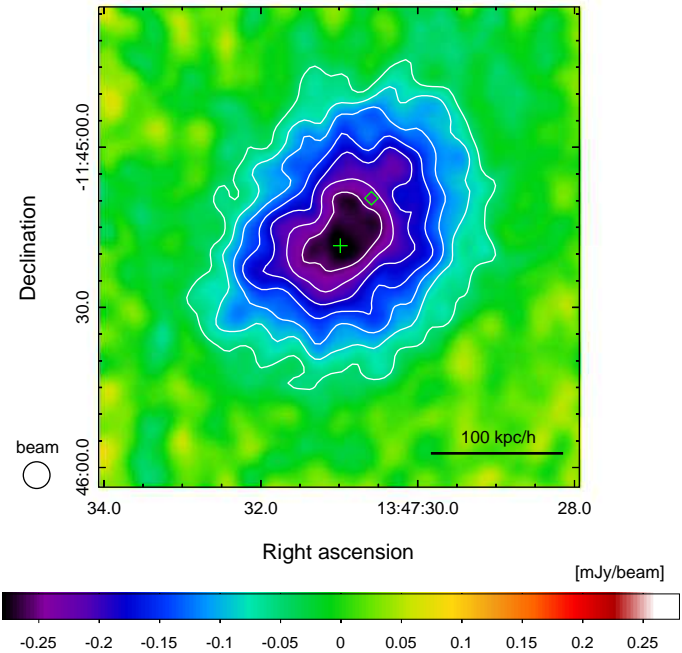} 
 \end{center}
\caption{Deconvolved map of RX J1347.5--1145 by ALMA at the
  central frequency of 92 GHz smoothed to have a symmetrical beam with
  $5''$ FWHM.  Contours show 3, 5, 7, 9, 11, 13, 15$\sigma$
  statistical significance levels with $1\sigma = 0.017$ mJy/beam. The
  positions of the SZE peak and the subtracted central AGN are marked
  by a cross and a diamond, respectively. }
\label{fig-sz}
\end{figure}

Figure \ref{fig-sz} shows the deconvolved map smoothed to an effective
beam size of $5''$ FWHM. Asymmetry of the synthesized beam has been
corrected, to study morphology of the emission.  Applying the same
smoothing to the difference map shown in the right panel of Figure
\ref{fig-dirty2}, we measure the rms noise level of 0.017 mJy/beam
over a diameter of $90''$ around the field center.  As described in
Section \ref{sec-source}, the rms values are nearly constant within
this diameter. We hence adopt 0.017 mJy/beam when denoting the
statistical significance within the same region of the smoothed image.

\begin{figure}
 \begin{center}
  \includegraphics[width=9cm]{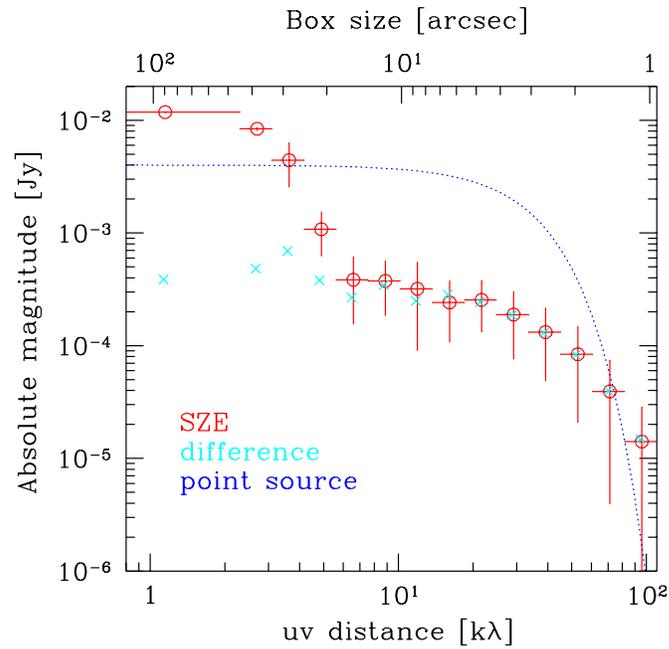} 
 \end{center}
\caption{Absolute magnitude of the Fourier transform of the
  deconvolved SZE map (circles) and the difference map in the right
  panel of Figure \ref{fig-dirty2} (crosses). Vertical error bars
  indicate standard deviations of the absolute magnitude in each
  bin. Also plotted for reference is the magnitude for a 4 mJy point
  source convolved with a Gaussian beam with the average FWHM of
  $3.2''$ (dashed line). The top axis shows the box size over which
  the data are sampled on the image. For clarity, symbols have been
  slightly shifted horizontally.}
\label{fig-uvprof}
\end{figure}

To examine characteristics of the signal and noise, we plot in Figure
\ref{fig-uvprof} the absolute magnitude (root sum square of real and
imaginary parts) of the Fourier transform of the deconvolved SZE map
and the difference map. Note that the plotted quantities are different
from observed visibilities. The Fourier transform was applied to
the unsmoothed maps over a square box of $90''\times 90''$ around the
peak of the decrement. The deconvolved map is dominated by an extended
signal over noise at $\gtrsim 15''$. As mentioned in Section
\ref{sec-deconv} and will be tested further using simulations in
Section \ref{sec-sim}, the extended signal up to the spatial scale of
about $40''$ is recovered in the current deconvolved map, whereas the
flux at a larger scale is lost owing to lack of the shorter spacing
data.

In Figure \ref{fig-sz}, the decrement signal from the SZE is detected
at more than 15 $\sigma$ statistical significance with $5''$
resolution; its peak position
(\timeform{13h47m31.02s},\timeform{-11d45m18.4})
is at $11''$ to the south-east from the central AGN.  A separation of
$11''$ is more than 8$\sigma$ away for the synthesized beam of ALMA
and the uncertainty of the AGN flux does not affect the intensity of
the south-east peak nor overall morphology of the image. The measured
intensities at the south-east peak and at the AGN position are $-0.286
\pm 0.017 \pm 0.017 $ mJy/beam and $-0.234 \pm 0.017 \pm 0.12$
mJy/beam, respectively (quoted systematic errors are from the flux
calibration and the source subtraction).  There is also a weak local
peak with $-0.221 \pm 0.017 \pm 0.013 $ mJy/beam at $7''$ to the
north-west of the central AGN.  The measured flux within a radius of
$40''$ from the south-east peak is $-12.4 \pm 0.2 \pm 0.8$ mJy. The
mean signal within the annulus at $40''-45''$ from the south-east peak
is consistent with zero and $0.0004 \pm 0.0010$ mJy/beam.

\begin{figure}
 \begin{center}
  \includegraphics[width=9cm]{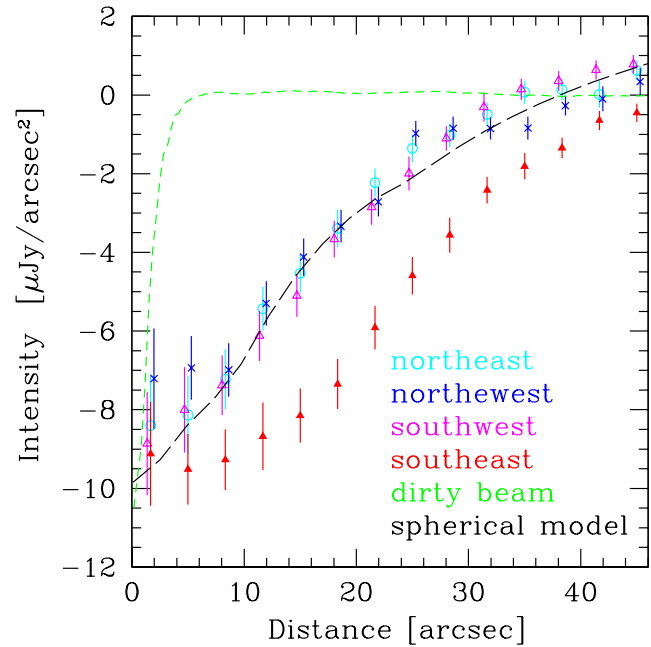} 
 \end{center}
\caption{Azimuthal average of the measured intensity as a function of
  the distance from the central AGN position in four quadrants;
  northeast (open circles), northwest (crosses), southwest (open
  triangles) south east (filled triangles). Error bars indicate
  statistical and systematic errors of the mean in each bin. For
  clarity, symbols have been slightly shifted horizontally. For
  reference, the short dashed line shows the azimuthally averaged
  shape of the dirty beam normalized to have the peak value of 0.12
  mJy/beam in the unsmoothed map.  The long dashed line shows the
  shape of a spherical model profile from the Chandra X-ray data
  excluding the south-east quadrant with the zero-brightness level
  taken at the distance of $38''$ from the central AGN; the model
  profile varies as $\propto h^{-1/2}$ and $h=0.7$ is assumed in the
  figure.}
\label{fig-szprof}
\end{figure}

We present in Figure \ref{fig-szprof} azimuthally averaged intensity
profiles as a function of the distance from the central AGN
position. Unsmoothed pixel data are binned and the statistical error
of the mean in each bin is computed by
\begin{eqnarray}
\sigma_{\rm bin} = \frac{1}{\sqrt{N_{\rm beam}}}\max(\sigma_{\rm std},
\sigma_{\rm map}),
\label{eq-bin}
\end{eqnarray}
where $\sigma_{\rm std}$ is the standard deviation of the pixel data
in the bin, $\sigma_{\rm map}$ is the map noise, and $N_{\rm beam}$ is
the number of independent beams in the bin (we use $N_{\rm beam}$
instead of the number of pixels, because the pixel values in a
deconvolved image are correlated within the beam area).  The
calibration error of 6\% is then added in quadrature to the error in
each bin.  To illustrate the impact of the systematic error of the
subtracted AGN flux, we plot in the same figure the azimuthally
averaged shape of the dirty beam (the point spread function before
sidelobes are corrected) normalized to have the peak value
corresponding to 0.12 mJy/beam in the unsmoothed map. Only the central
bin may potentially be affected by the error of the subtracted AGN
flux.

Figure \ref{fig-szprof} shows that the intensity in the south-east
quadrant is significantly stronger, whereas extended signals are
present in all the quadrants. The average signal excluding the
south-east quadrant drops to zero at $38''$ from the central AGN.  For
reference, we also plot a spherically symmetric SZE model based on the
Chandra X-ray data of this cluster excluding the south-east
quadrant. Spatially resolved and deprojected electron temperatures are
taken from Table 7 of \citet{Ota08} with the relativistic correction
of the SZE by \citet{Nozawa05} at 92~GHz; the radial electron density
profile is modeled by a sum of two $\beta$-models \citep{Cavaliere78},
$n_{\rm e}(r) = n_{{\rm e1}}[1+(r/r_{\rm c1})^2]^{-3\beta/2} + n_{{\rm
    e2}}[1+(r/r_{\rm c2})^2]^{-3\beta/2}$, with the parameters
obtained from fitting the X-ray brightness at 0.4--7.0 keV, $n_{{\rm
    e1}}= (0.241 \pm 0.007) ~h^{1/2}$cm$^{-3}$, $r_{\rm c1}=(9.14 \pm
0.74 ) ~h^{-1}$kpc, $n_{{\rm e2}}=(0.055 \pm 0.008)
~h^{1/2}$cm$^{-3}$, $r_{\rm c2}=(36.6 \pm 2.7) ~h^{-1}$kpc, and
$\beta=0.568 \pm 0.004$; the projected SZE intensity is convolved with
the synthesized beam of ALMA; and the zero-brightness level is taken
at $38''$ from the central AGN (the AGN lies less than $1''$ from the
X-ray peak).  Note that the spherical model SZE profile varies as
$\propto h^{-1/2}$ and $h=0.7$ is assumed for definiteness in Figure
\ref{fig-szprof}.  The signal in the south-east quadrant is clearly
stronger than the other quadrants. The average intensity profile
excluding the south-east quadrant is in broad agreement with the
spherical model, although it tends to be weaker by $\sim 10 \%$.  The
difference may be ascribed to the kinematic SZE, asphericity of the
cluster, the missing flux of the ALMA map, the calibration error of
ALMA (Section \ref{sec-obs}) or Chandra \citep{Reese10}, a higher
value of $h$, or any combination thereof.  Regarding the first
possibility, the peculiar velocity of this cluster $v_{\rm
  pec}=-1040^{+870}_{-840}$ km/s reported by \citet{Sayers16} reduces
the observed SZE signal at 92~GHz by $8.8 ^{+7.1}_{-7.4}~\%$ for the
mean electron temperature of $kT_{\rm e}=13$~keV \citep{Ota08}.  We
will discuss the second and the third possibilities in detail in
Sections \ref{sec-x} and \ref{sec-sim}.

The above results demonstrate ALMA's powerful capability of detecting
an extended SZE signal.  They also confirm, with much improved spatial
resolution and sensitivity, the previous findings
\citep{Komatsu01,Allen02} of a substructure in the south-east quadrant
of this cluster and identify its location regardless of the presence
of the central AGN. Their implications including detailed comparison
with the X-ray data as well as with the previous SZE observations of
this cluster are presented in the following sections.

\section{Interpretation and Implications}
\label{sec-implication}

\subsection{X-ray Data of RX~J1347.5--1145}
\label{sec-x}

To perform detailed comparison with the ALMA results (Section
\ref{sec-sz}) as well as to construct a realistic model for imaging
simulations (Section \ref{sec-sim}), we extracted 6 data sets of
RX~J1347.5--1145 taken by ACIS-S (ObsID 506 and 507) and ACIS-I (ObsID
3592, 13516, 13999, and 14407) on board Chandra X-ray
observatory. After excluding the periods with high background rates,
the total net exposure time amounts to 233.8 ks.  The data were
processed using CIAO version 4.7 \citep{Fruscione06} and the
Calibration database (CALDB) version 4.6.9. Given that the cluster is
highly compact compared to the field-of-view of ACIS, we estimated the
backgrounds from the off-center region at $2.5' - 3.5'$ away from the
X-ray peak of this cluster, where the intracluster emission is
negligible. Exposure-corrected and background-subtracted data at
0.4--7.0 keV were used throughout our analysis. Spectral fitting was
done with XSPEC version 12.9.0o \citep{Arnaud96}, assuming that the
intracluster plasma is in collisional ionization equilibrium and the
metal abundance ratio is that of \citet{Anders89}. The source redshift
and Galactic hydrogen column density were fixed at $z=0.451$ and
$N_{\rm H}=4.6\times 10^{20}$ cm$^{-2}$ \citep{Kalberla05},
respectively.

We performed X-ray thermodynamic mapping using the contour binning
algorithm \citep{Sanders06} to define sub-regions with nearly equal
photon counts. Fitting a spectrum in each sub-region defined by this
algorithm yields projected temperature, metallicity, and spectral
normalization factor. From these quantities, ``pseudo'' electron
density and pressure were computed assuming that the gas is
distributed uniformly over the line-of-sight distance of $L$.  These
are not real electron density or pressure maps, as we did not perform
any deprojection to estimate a three-dimensional distribution. The
pseudo electron density map is essentially the square-root of the
projected X-ray intensity ($\propto \int n_e^2 dl)$, while the pseudo
pressure map is a product of the pseudo density and the projected
X-ray temperature. The absolute values of the density and the pressure
are arbitrary (both varies as $L^{-1/2}$) and only morphology of the
maps is to be compared with the SZE.  We adopt the S/N threshold of 83
($\sim 7000$ counts) to obtain typical statistical errors $5\%$ for
the density and $10\%$ for the temperature and the pressure.

\begin{figure}
 \begin{center}
  \includegraphics[width=8cm]{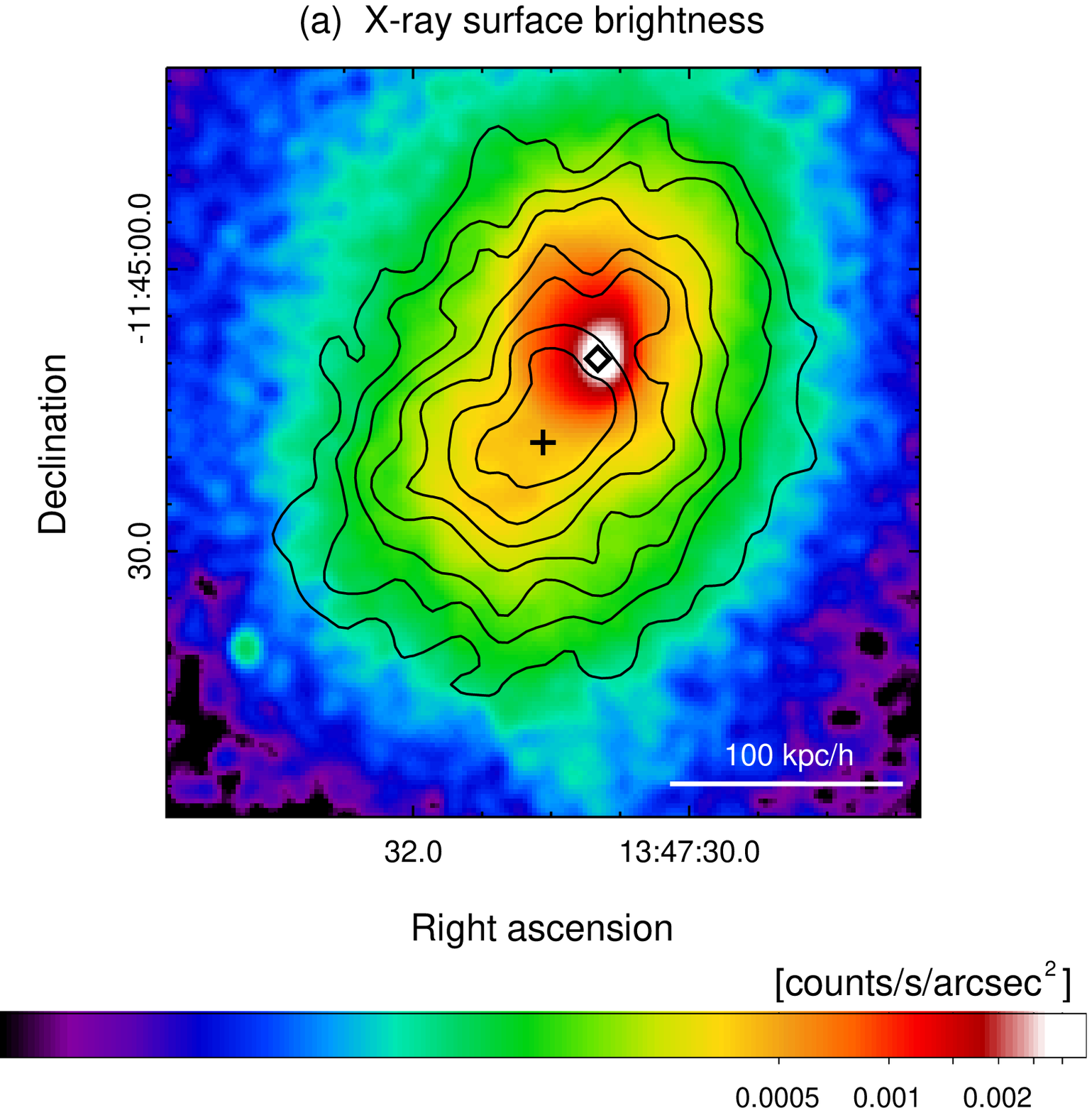} 
  \includegraphics[width=8cm]{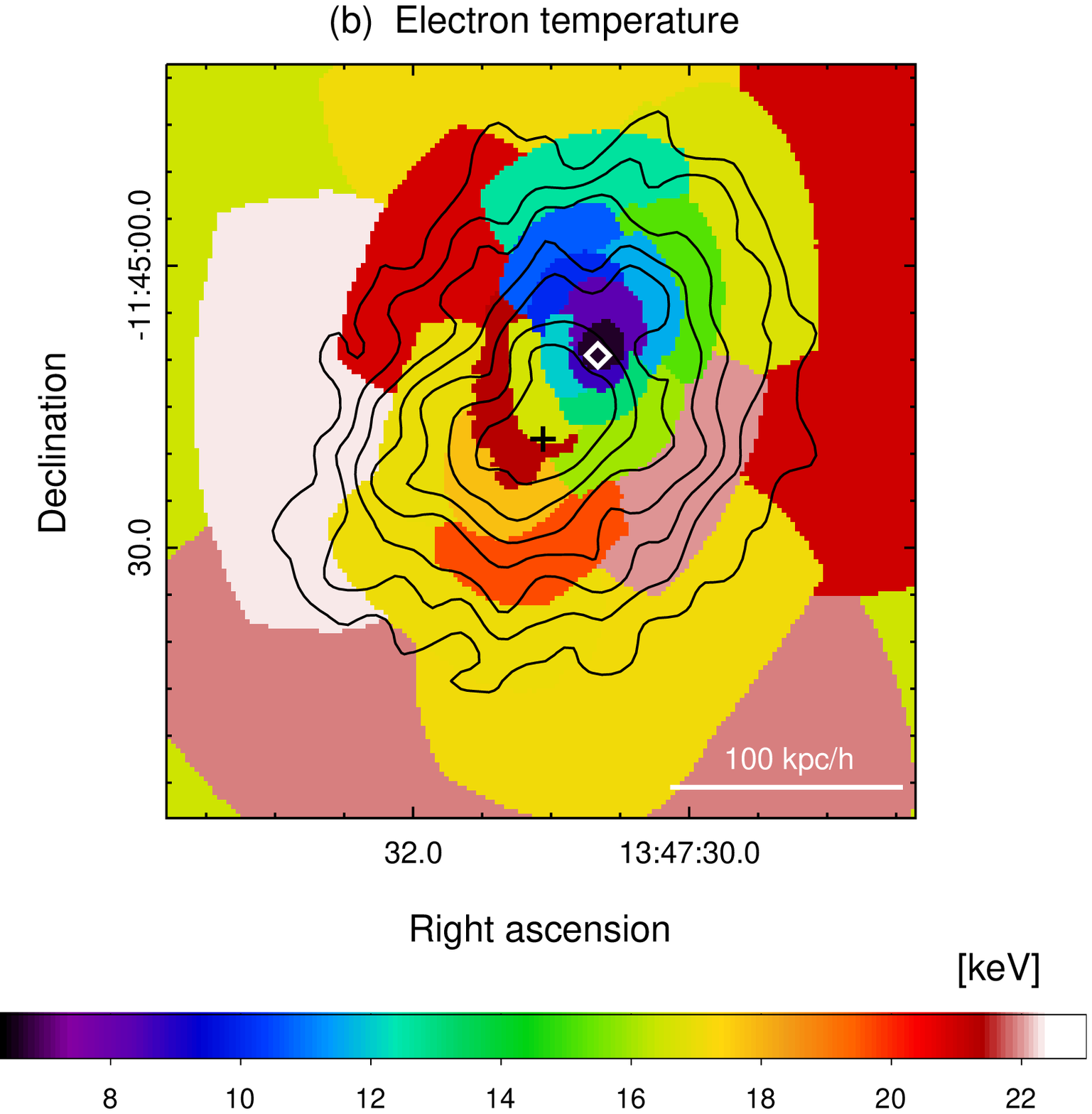} 
  \includegraphics[width=8cm]{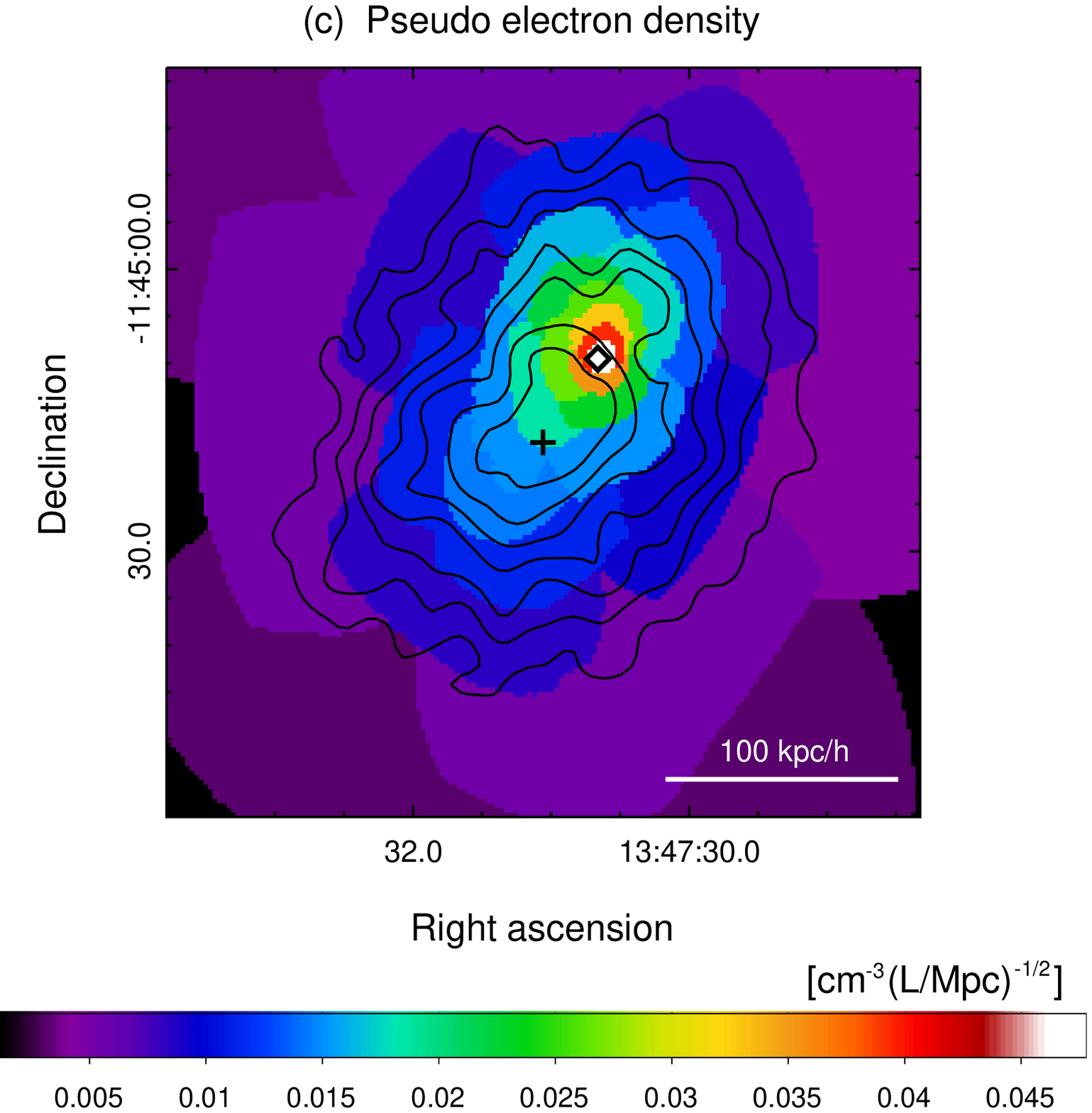} 
  \includegraphics[width=8cm]{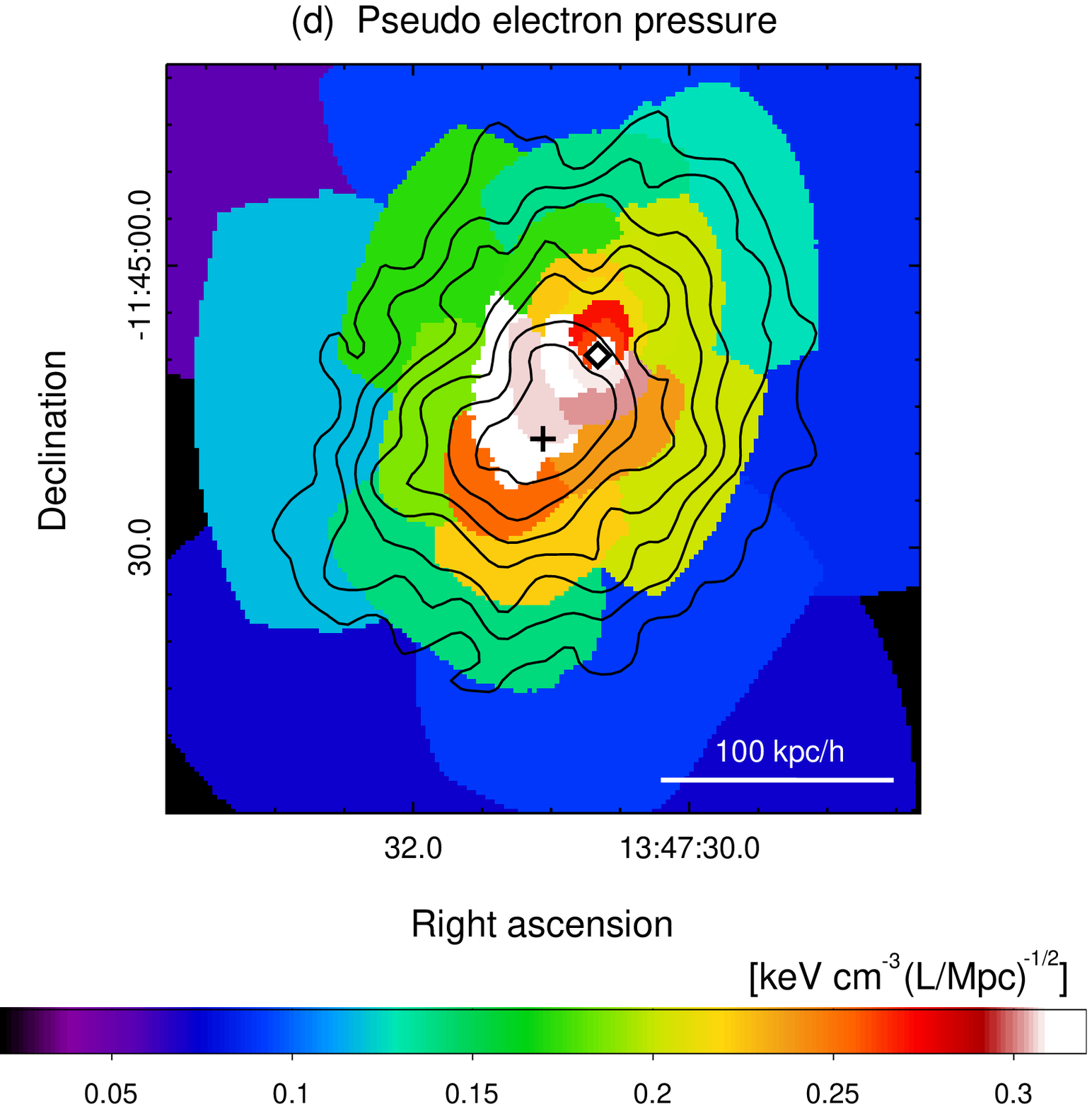} 
 \end{center}
\caption{Comparison of the ALMA SZE data and the Chandra X-ray images
  of RX~J1347.5--1145.  In all panels, contours indicate the
  significance levels of the ALMA SZE map plotted in Figure
  \ref{fig-sz}, and a cross and a diamond indicate the positions of
  the SZE peak and the central AGN, respectively.  (a) The X-ray
  surface brightness at 0.4--7.0 keV in counts/s/arcsec$^2$, smoothed
  by a Gaussian kernel with $2.3''$ FWHM. (b) Projected X-ray
  spectroscopic temperature in keV, based on the contour binning
  algorithm \citep{Sanders06} with the S/N threshold of 83 (7000
  counts) per region. (c) Pseudo electron density map in cm$^{-3}
  (L/{\rm Mpc})^{-1/2}$ assuming a uniform line-of-sight depth of
  $L$. (d) Pseudo electron pressure map in keV cm$^{-3} (L/{\rm
    Mpc})^{-1/2}$, a product of quantities plotted in panels (b) and
  (c).}
\label{fig-xray}
\end{figure}

Figure \ref{fig-xray} compares the X-ray measured quantities with the
ALMA SZE contours.  The pseudo pressure map (panel d) shows a
reasonable agreement with the ALMA SZE map, including the south-east
peak position and elongated structure of the emission. The projected
mean temperature (panel b) exceeds 20 keV around the SZE peak in
accord with the results of \citet{Kitayama04} and \citet{Ota08}.
While departure from spherical symmetry is indicated in all the maps,
disturbance is more obvious in the SZE, temperature, and pressure maps
than in the X-ray brightness or density maps. Being proportional to
the density squared, the X-ray brightness tends to peak sharply within
the central cool core. The spatial resolution of the ALMA map ($20
~h^{-1}$kpc) allows us to clearly separate the cool core and the
south-east pressure peak.

The position of the central AGN identified in the ALMA data agrees
with the X-ray peak to within $1''$. The central AGN is faint in
X-rays with an upper limit ($1\sigma$) on the 2--10 keV luminosity of
$4.38 \times 10^{42}$ ergs/s.

\subsection{Imaging Simulations: How Much Flux is Missing 
on Large Scales?}
\label{sec-sim}

Given good agreement between the ALMA map and the X-ray inferred
pressure map derived in the previous section, we performed imaging
simulations using the latter as a realistic input model to quantify
the degree of missing flux for RX J1347.5--1145.
The absolute value of the pressure was normalized so that the peak
value corresponds to the Compton $y$-parameter of $y_{\rm
  peak}=8\times10^{-4}$ as inferred from the previous SZE measurements
of this cluster \citep{Komatsu01,Kitayama04}. A relativistic
correction to the SZE intensity by \citet{Nozawa05} was applied
adopting the projected temperature shown in Figure \ref{fig-xray}(b)
at each sky position.

We created model images separately at four spectral windows centered
at 85, 87, 97, and 99 GHz with an effective bandwidth of 1.875 GHz
each. The pointing directions, the array configuration, the hour
angle, the total effective integration time, and the average
perceptive water vapour were set to match those of each executing
block of real observations. Visibility data were then produced using
the CASA task {\it simobserve}. Three sets of the data were produced
and used for different purposes; (1) visibility without signal but
includes instrumental and atmospheric noise expected for ALMA in each
spectral window, (2) visibility without noise but includes signal, and
(3) visibility with both noise and signal. The rms levels of dirty
images created from the first set of visibility were consistent with
the values given in Section \ref{sec-source}. The second and the third
sets of visibility were deconvolved in the same way as the real data
as described in Section \ref{sec-deconv}. In the following, the
simulation results are compared with the input signal at the central 
frequency 92 GHz, to take account of any bias arising from spectral
bandpass of the observations and the data analysis.

\begin{figure}
 \begin{center}
  \includegraphics[height=6.1cm]{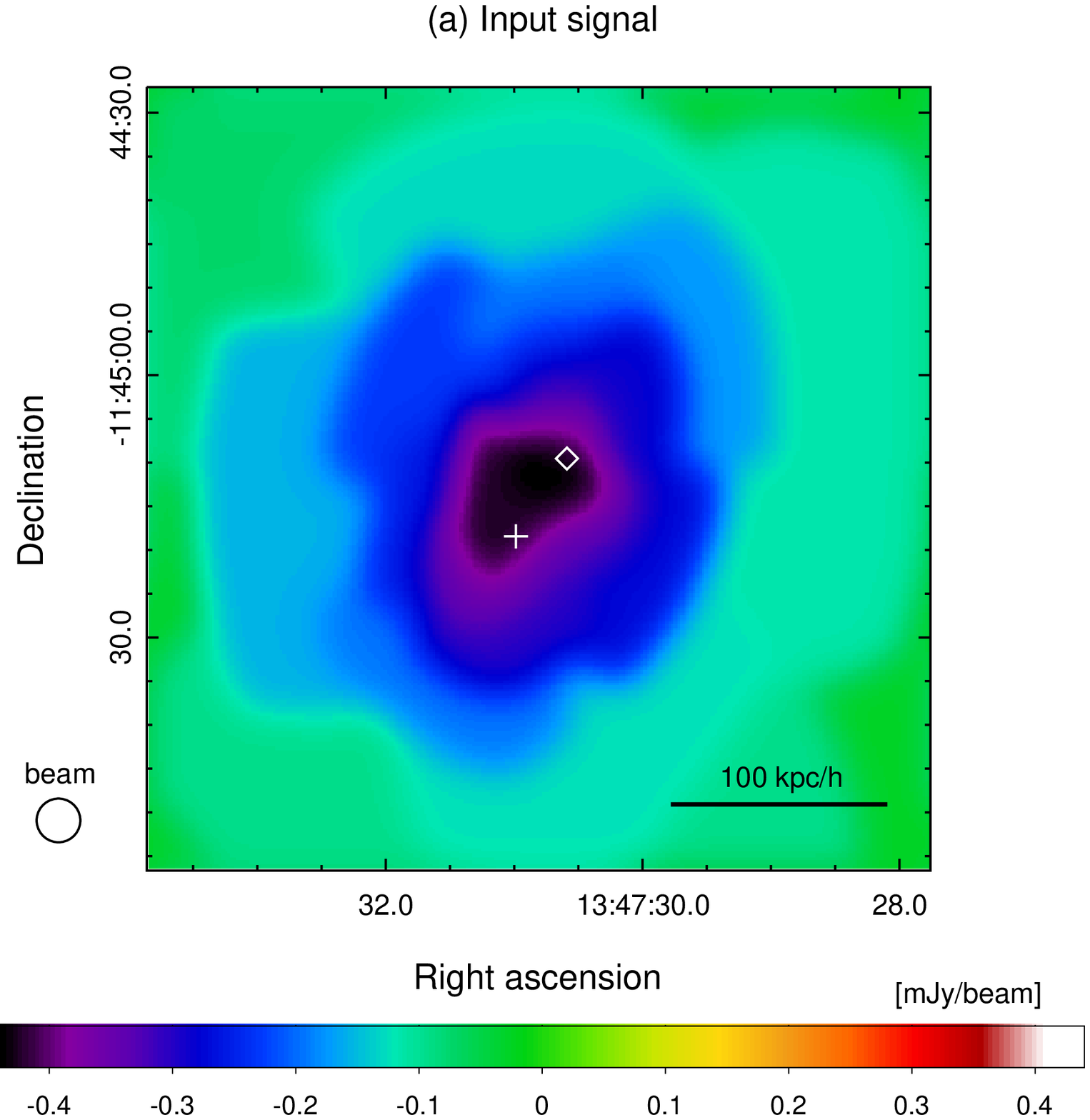} 
  \includegraphics[height=6.1cm]{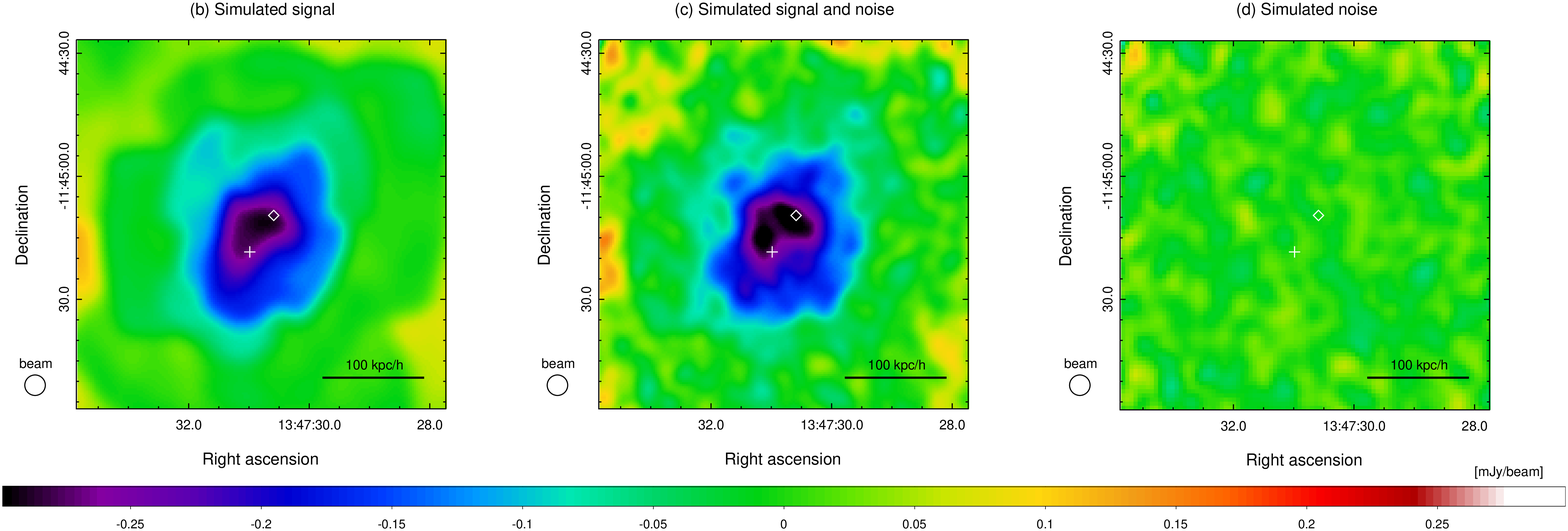} 
 \end{center}
\caption{Mock ALMA SZE maps of RX J1347.5--1145 with $y_{\rm
    peak}=8\times 10^{-4}$. (a) Input map at 92 GHz. (b) Simulated map
  with signal only. (c) Simulated map with both signal and noise. (d)
  Simulated map with noise alone; the rms value over the plotted
  region is $20~\mu$Jy/beam.  All the maps have been smoothed to the
  resolution of $5''$ FWHM and have the size $90''\times 90''$
  centered at the emission peak of the input map. For reference, the
  positions of the central AGN and the SZE peak in the {\it observed}
  ALMA map are marked by a diamond and a cross, respectively.}
\label{fig-simmap1}
\end{figure}
\begin{figure}
 \begin{center}
  \includegraphics[width=8cm]{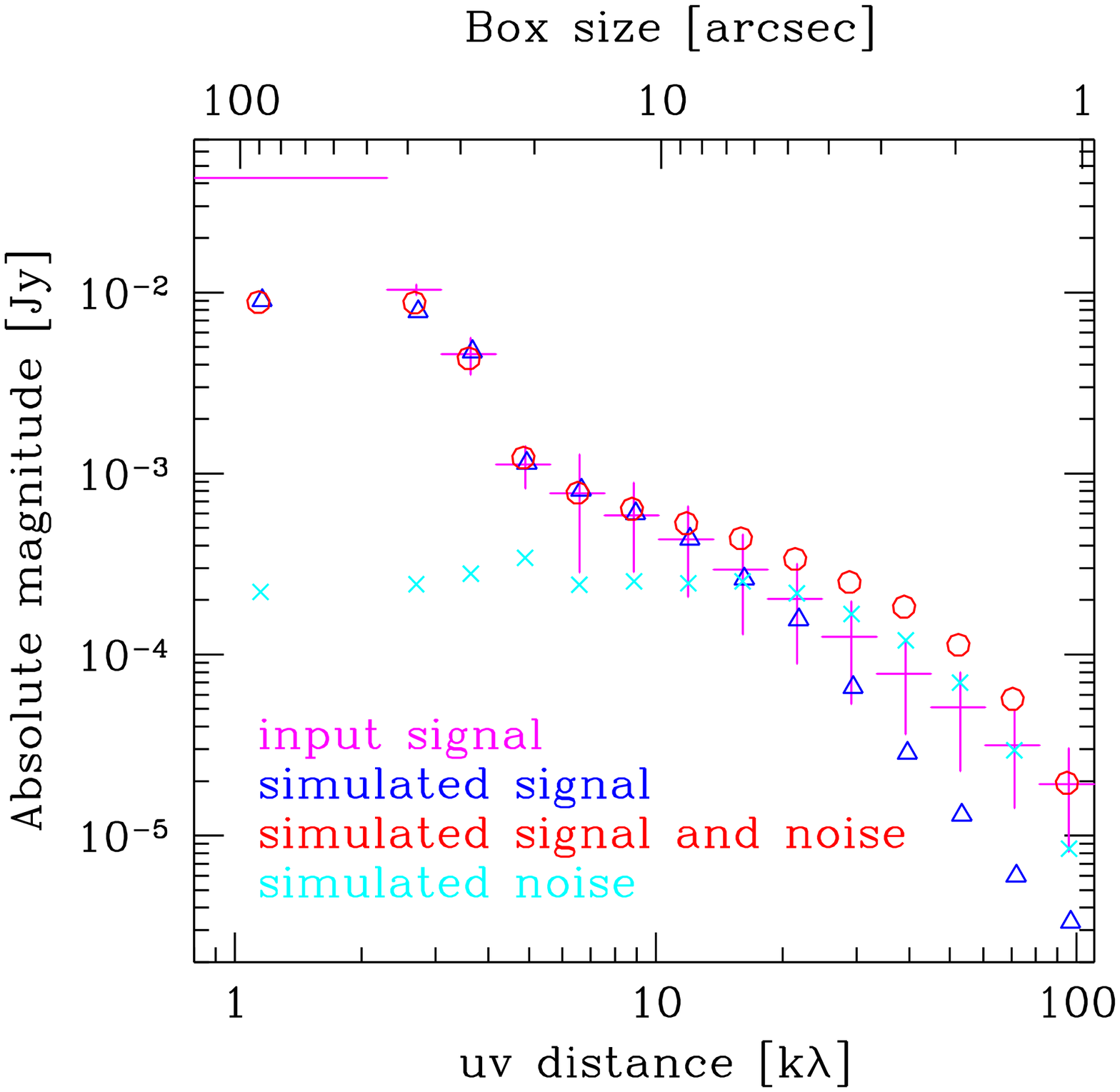} 
  \includegraphics[width=8cm]{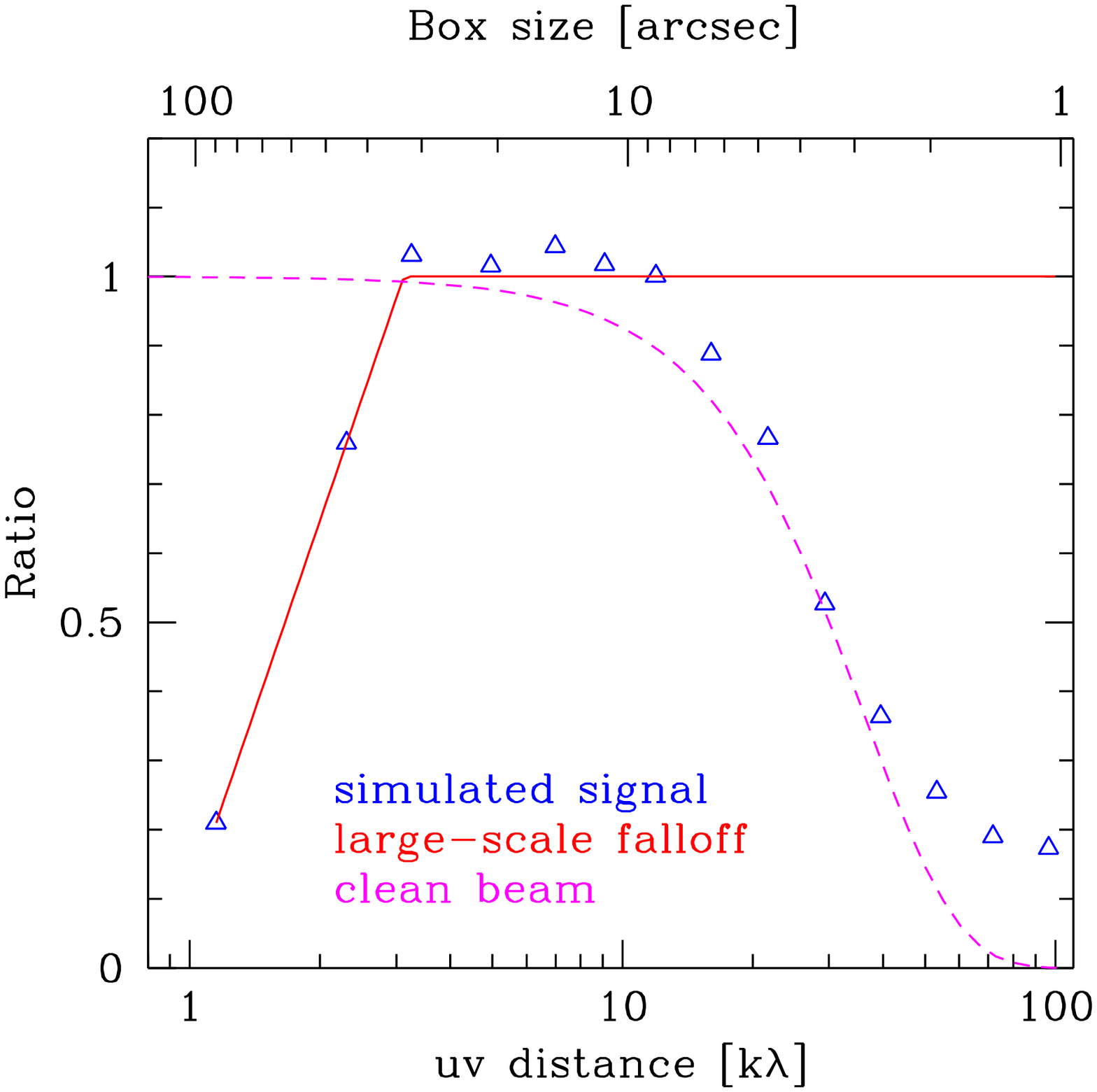} 
 \end{center}
\caption{Fourier transform of mock ALMA maps. {\it Left:} Absolute
  magnitude of the input map at 92 GHz before observed by ALMA (error
  bars), simulated ALMA map with signal only (triangles), simulated
  ALMA map with both signal and noise (circles), and simulated ALMA
  map with noise alone (crosses). Vertical error bars indicate
  standard deviations of the absolute magnitude in each bin. For
  clarity, they are plotted only for the input map and those of the
  simulated map with signal and noise are similar to those of the
  observed map plotted in Figure \ref{fig-uvprof}. Symbols have been
  slightly shifted horizontally for display {\it Right:} Ratio of the
  absolute magnitude of the Fourier transform of the simulated signal
  without noise to that of the input map (triangles), the factor to
  model a falloff at large spatial scales (solid line), and a Gaussian
  clean beam to account for finite spatial resolution (dashed
  line). The top axis shows the box size over which the data are
  sampled on the image.}
\label{fig-uvprof_sim}
\end{figure}

Figure \ref{fig-simmap1} compares the simulated ALMA maps to the
  input; note that the range of the color scale of panel (a) is wider
  than that of panels (b)--(d). The amplitude of the Fourier transform
  of each map is plotted in Figure \ref{fig-uvprof_sim}. The
  simulated signal and noise both show similar magnitude to the real
  data plotted in Figure \ref{fig-uvprof}.  A large fraction ($\sim
  80\%$) of flux is lost at the smallest $uv$ distance, whereas
  reconstruction becomes accurate at $2.5 - 20 $~k$\lambda$,
  corresponding to the spatial size of $40'' - 5''$.  At larger
  $uv$ distances, reconstruction is limited by noise and finite
  spatial resolution. These results suggest that the amount of the
  missing flux is controlled primarily by the $uv$ coverage and the
  deconvolution method.  An azimuthally-averaged amplitude at a fixed
  $uv$ distance intrinsically has a large dispersion owing to
  asymmetry of the emission. In the following, we present two
  different approaches to model the amount of the missing flux.

\begin{figure}
 \begin{center}
  \includegraphics[width=16.5cm]{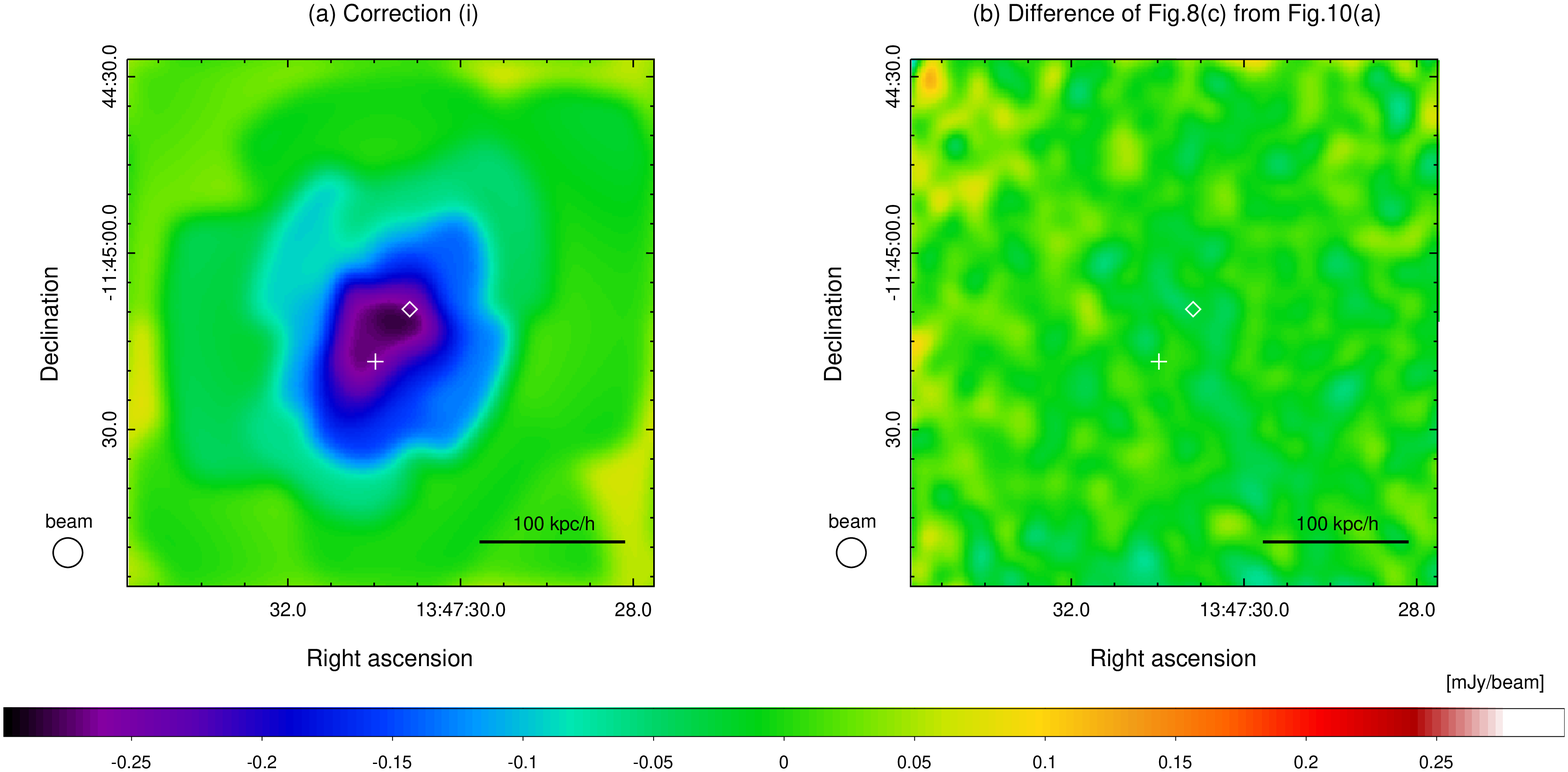} 
  \includegraphics[width=16.5cm]{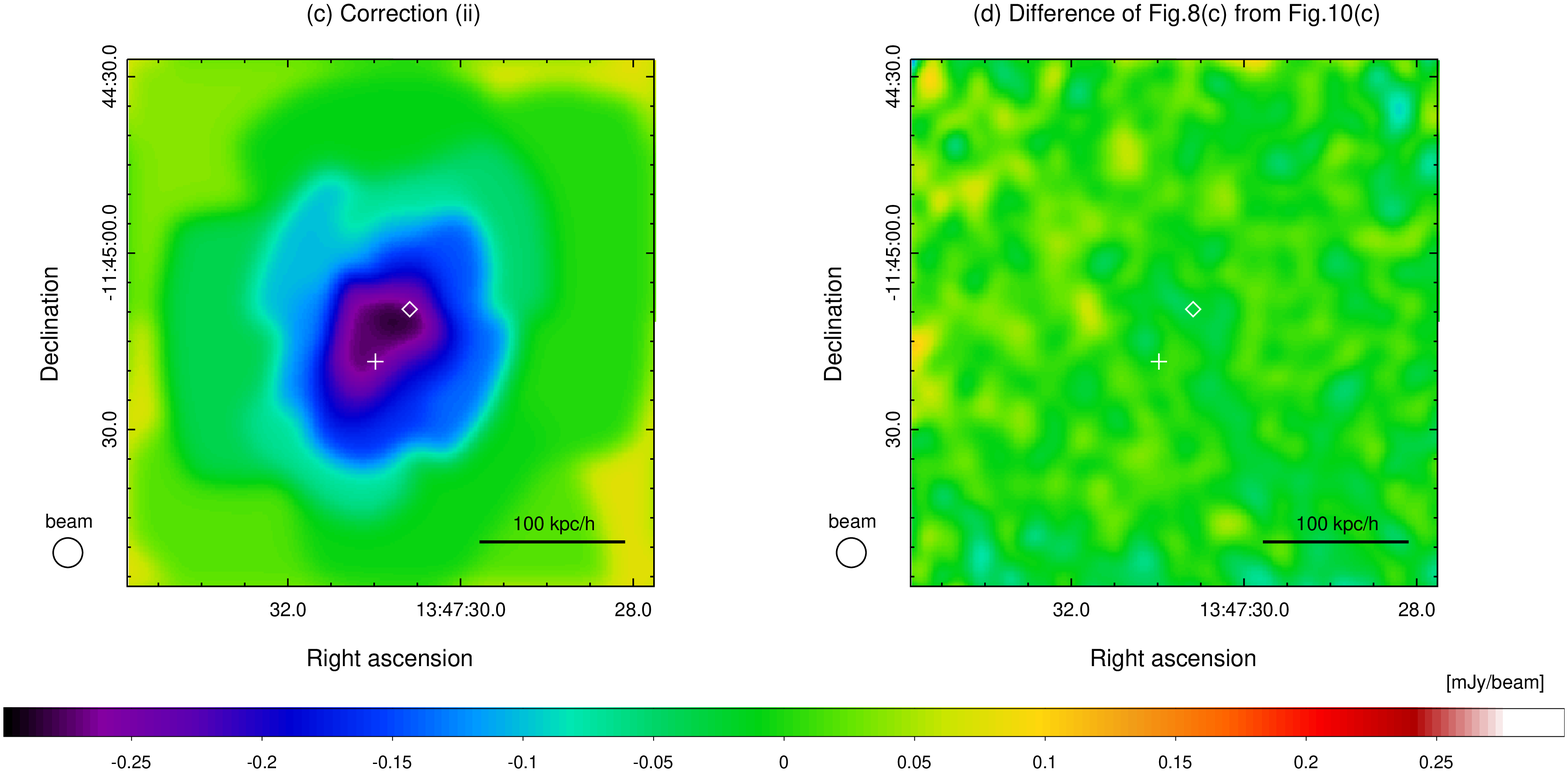} 
 \end{center}
\caption{Model SZE maps of RX J1347.5--1145 with $y_{\rm peak}=8\times
  10^{-4}$ corrected for the missing flux of the ALMA
  observations. (a) Input map to which correction (i) is applied. (b)
  The difference of Figure\ref{fig-simmap1}(c) from
  Figure\ref{fig-simmap2}(a); the rms value of the map is
  $22~\mu$Jy/beam. (c) Input model to which correction (ii) is applied
  using equation (\ref{eq-conv}). (d) The difference of
  Figure\ref{fig-simmap1}(c) from Figure\ref{fig-simmap2}(c); the rms
  value of the map is $22~\mu$Jy/beam.  All the maps have been
  smoothed to the resolution of $5''$ FWHM and have the size
  $90''\times 90''$ centered at the emission peak of the input
  map. For reference, the positions of the central AGN and the SZE
  peak in the {\it observed} ALMA map are marked by a diamond and a
  cross, respectively. The range of the color scale is the same as
  Figures \ref{fig-simmap1}(b)--(d). }
\label{fig-simmap2}
\end{figure}

The first approach (hereafter correction (i)) is to correct the
intrinsic signal in Fourier space.  The right panel of Figure
\ref{fig-uvprof_sim} shows the ratio of the amplitude of simulated
signal (without noise) to that of the input map.  Apart from the
dispersion arising from asymmetry of the emission, the average ratio
is well described by a product of a large-scale fall-off factor (solid
line) obtained by interpolating the points at the shortest $uv$
distances, and a Gaussian clean beam (dashed line) that accounts for a
finite angular resolution of the image.
A correction is then applied by multiplying these two factors to the
Fourier transform of the input model at 92~GHz; in practice, the
latter factor was applied equivalently as a convolution with a
Gaussian kernel in real space.  Figures \ref{fig-simmap2}(a) and (b)
show that correction (i) reproduces well the simulated map with
both signal and noise; the residual is consistent with noise plotted
in Figure \ref{fig-simmap1}(d).

\begin{figure}
 \begin{center}
  \includegraphics[width=8cm]{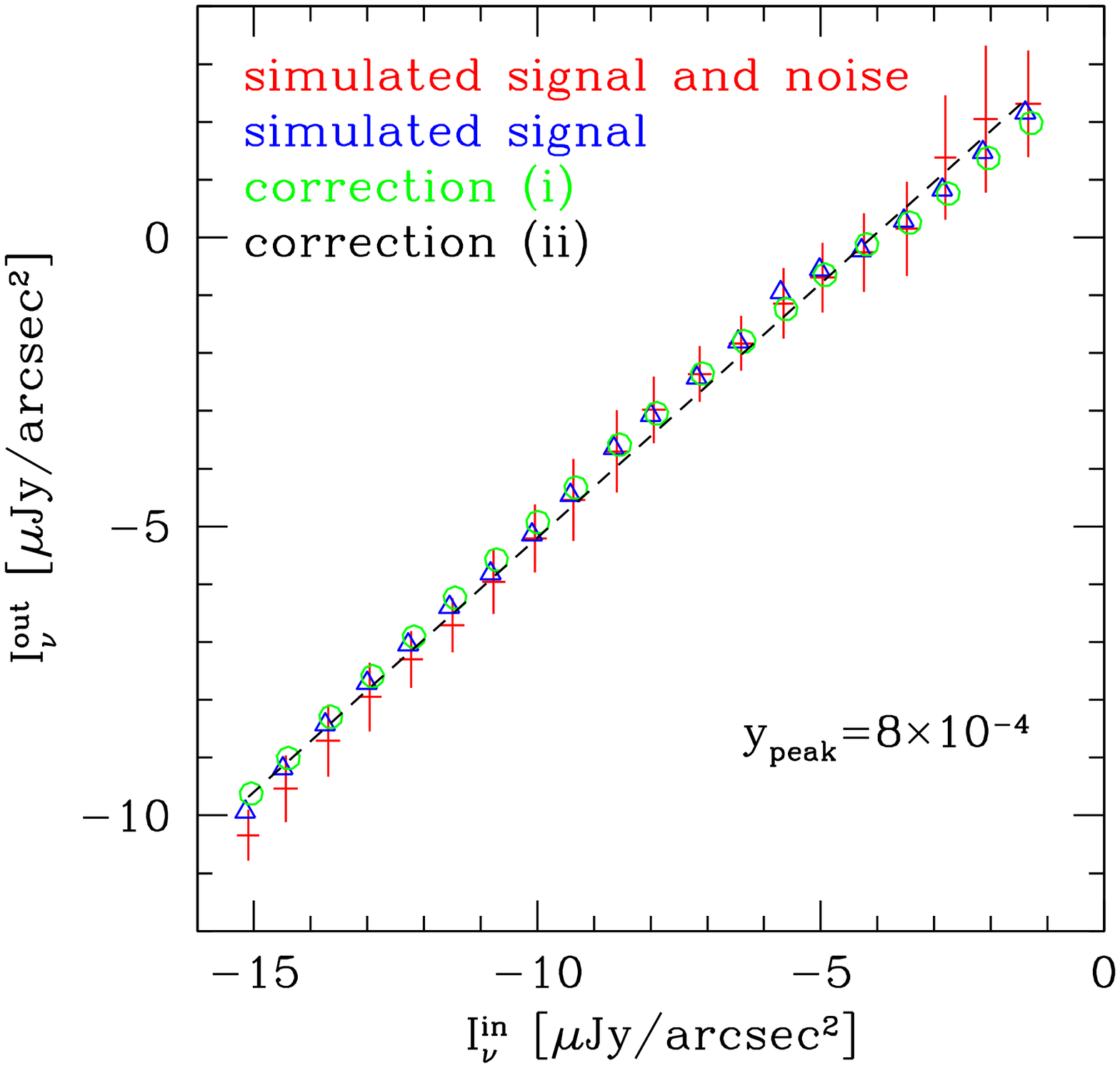} 
  \includegraphics[width=8cm]{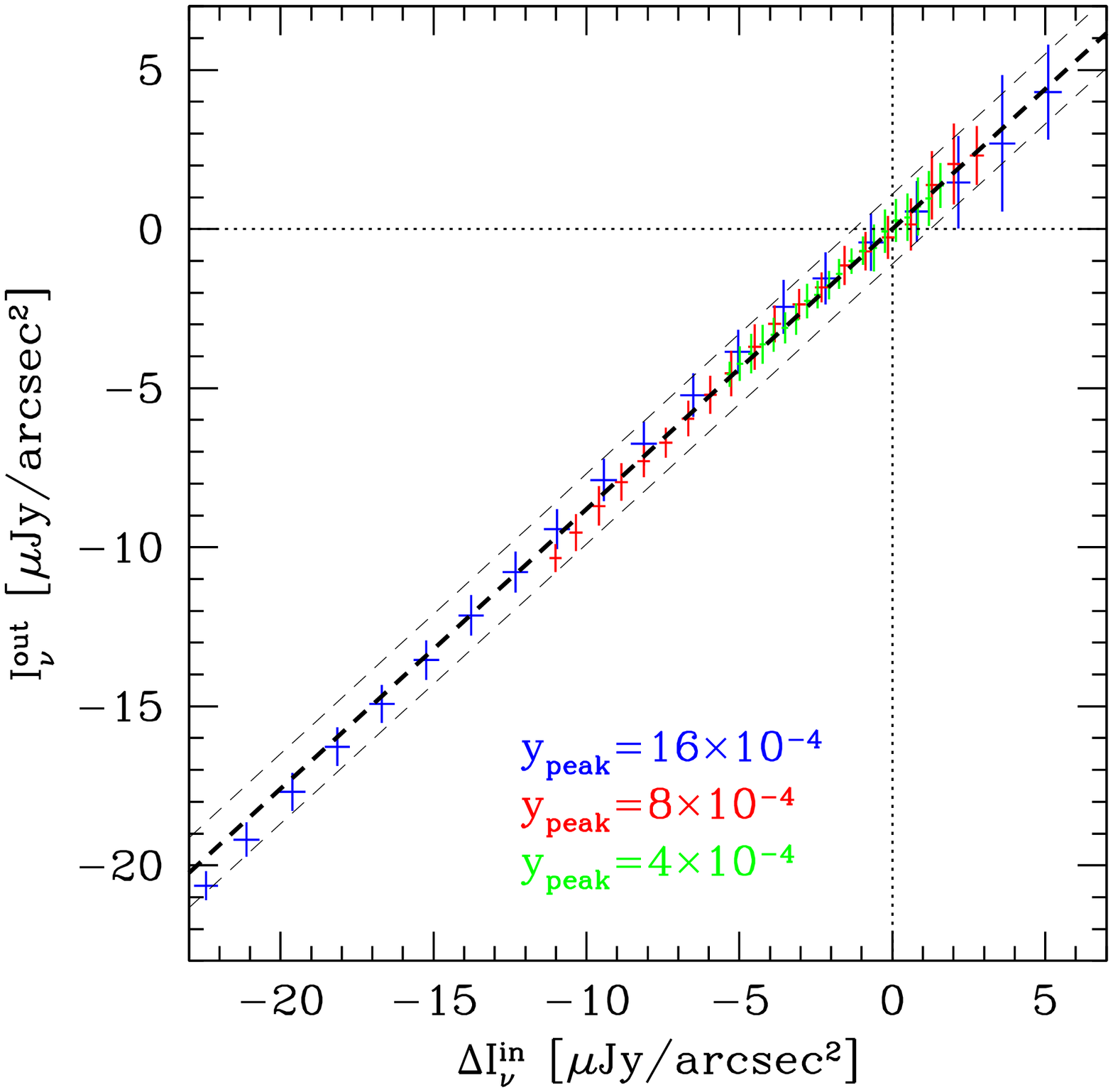} 
 \end{center}
\caption{Relation between the output intensity and the input intensity
  at 92~GHz from the simulations for RX J1347.5--1145.  Horizontal and
  vertical error bars denote standard deviations in each bin. For
  clarity, they are plotted only for the simulated map with both
  signal and noise. {\it Left:} Error bars indicate the simulation
  with both signal and noise for $y_{\rm peak}=8\times 10^{-4}$.  Also
  plotted are the simulation with signal only (triangles), the input
  model to which correction (i) is applied (circles), the input model
  to which correction (ii) is applied using equation (\ref{eq-conv})
  (dashed line).  For clarity, symbols have been slightly shifted
  horizontally. {\it Right:} Error bars indicate the simulation with
  both signal and noise for $y_{\rm peak}=16\times 10^{-4}$ (blue),
  $8\times 10^{-4}$ (red), and $4\times 10^{-4}$ (green), against the
  input intensity after a constant offset ($\Delta I_{\nu}^{\rm
    in}=+7.8,~+4.1,~+2.2~\mu$Jy/arcsec$^2$, respectively) is applied
  to give zero intercept. The thick dashed line shows the relation
  given by equation (\ref{eq-conv}) but taking $c_0=0$, and the thin
  dashed lines indicate the range of $1\sigma$ systematic
  uncertainties from it ($\Delta I_{\nu}^{\rm in}= \pm
  1.3~\mu$Jy/arcsec$^2$). }
\label{fig-conversion}
\end{figure}

The second approach (hereafter correction (ii)) relates the input and
output maps directly in real space.  Figure \ref{fig-conversion}(a)
compares the intensity of the output map from the simulation with both
signal and noise $I_{\nu}^{\rm out}$ to that of the input map
$I_{\nu}^{\rm in}$ (error bars), both smoothed to the beam size of
$5''$ FWHM. A set of data ($I_{\nu}^{\rm out}$, $I_{\nu}^{\rm in}$)
was created for each sky position and then binned in an ascending
order of $I_{\nu}^{\rm in}$; the error bars on $I_{\nu}^{\rm out}$ and
$I_{\nu}^{\rm in}$ in Figure \ref{fig-conversion} indicate standard
deviations in each bin.  For comparison, the same binning procedure was
repeated by replacing $I_{\nu}^{\rm out}$ with the intensity of the
simulated map with only signal (triangles) as well as the input map to
which correction (i) is applied (circles). A tight correlation is
found among these quantities. The dashed line in Figure
\ref{fig-conversion} indicates a linear approximation to this
correlation obtained from a fit to the {\it unbinned} set of
($I_{\nu}^{\rm out}$, $I_{\nu}^{\rm in}$):
\begin{eqnarray}
I_{\nu}^{\rm out} = c_1 I_{\nu}^{\rm in} + c_0
\label{eq-conv}
\end{eqnarray}
with $c_1=0.88$ and $c_0=3.6 ~\mu$Jy/arcsec$^2$.  The root-mean-square
deviation of the unbinned set of ($I_{\nu}^{\rm out}$, $I_{\nu}^{\rm
  in}$) from the best-fit relation is $\Delta I_{\nu}^{\rm in}=0.88
~\mu$Jy/arcsec$^2$; we will adopt $\sqrt{2}$ times this value ($\Delta
I_{\nu}^{\rm in}=1.3 ~\mu$Jy/arcsec$^2$) as the $1\sigma$ error of the
relation when a constant offset ($c_0$) is subtracted later in our
analysis.  Correcting the input model using equation (\ref{eq-conv})
also gives a good match to the simulated map with both signal and
noise (Figures \ref{fig-simmap2} c,d) and is consistent with the
results of correction (i) (Figure \ref{fig-conversion}a).

To test robustness of correction (ii), we repeated the simulations by
doubling or halving the signal of the entire input map to have $y_{\rm
  peak}=16 \times 10^{-4}$ or $4 \times 10^{-4}$, and adopting
different noise realizations.  A similar relation was found between
$I_{\nu}^{\rm out}$ and $I_{\nu}^{\rm in}$ but with different values
of $c_0$; they converge on a single relation within the range of
quoted uncertainties ($\Delta I_{\nu}^{\rm in}= \pm 1.3
~\mu$Jy/arcsec$^2$) once the offset of the zero brightness level is
removed as plotted in the right panel of Figure
\ref{fig-conversion}. A physical reason for self-similarity of the
relation is that multiplying the large-scale fall-off factor in
Fourier space is equivalent to subtracting the signal convolved with a
corresponding kernel function in real space, which leaves nearly a
constant fraction of the original signal on the deconvolved map. The
specific form of the relation depends on the source shape and the
observing configuration.

The above results imply that the intensity measured on the observed
ALMA map of RX J1347.5--1145 provides a reasonable representation of
{\it differential} values of the true intensity. The missing flux
primarily resides in a constant offset of the entire map. An
additional reduction factor of $12\%$ ($c_1$ in equation
~[\ref{eq-conv}]) and its uncertainty $\Delta I_{\nu}^{\rm in}= \pm
1.3 ~\mu$Jy/arcsec$^2$ (including the error of subtracting $c_0$ in
equation ~[\ref{eq-conv}]) should also be taken into account when the
measured intensity is converted to the intrinsic intensity using
correction (ii).

\subsection{Differential $y$-parameter map}
\label{sec-yparam}

\begin{figure}
 \begin{center}
  \includegraphics[height=10cm]{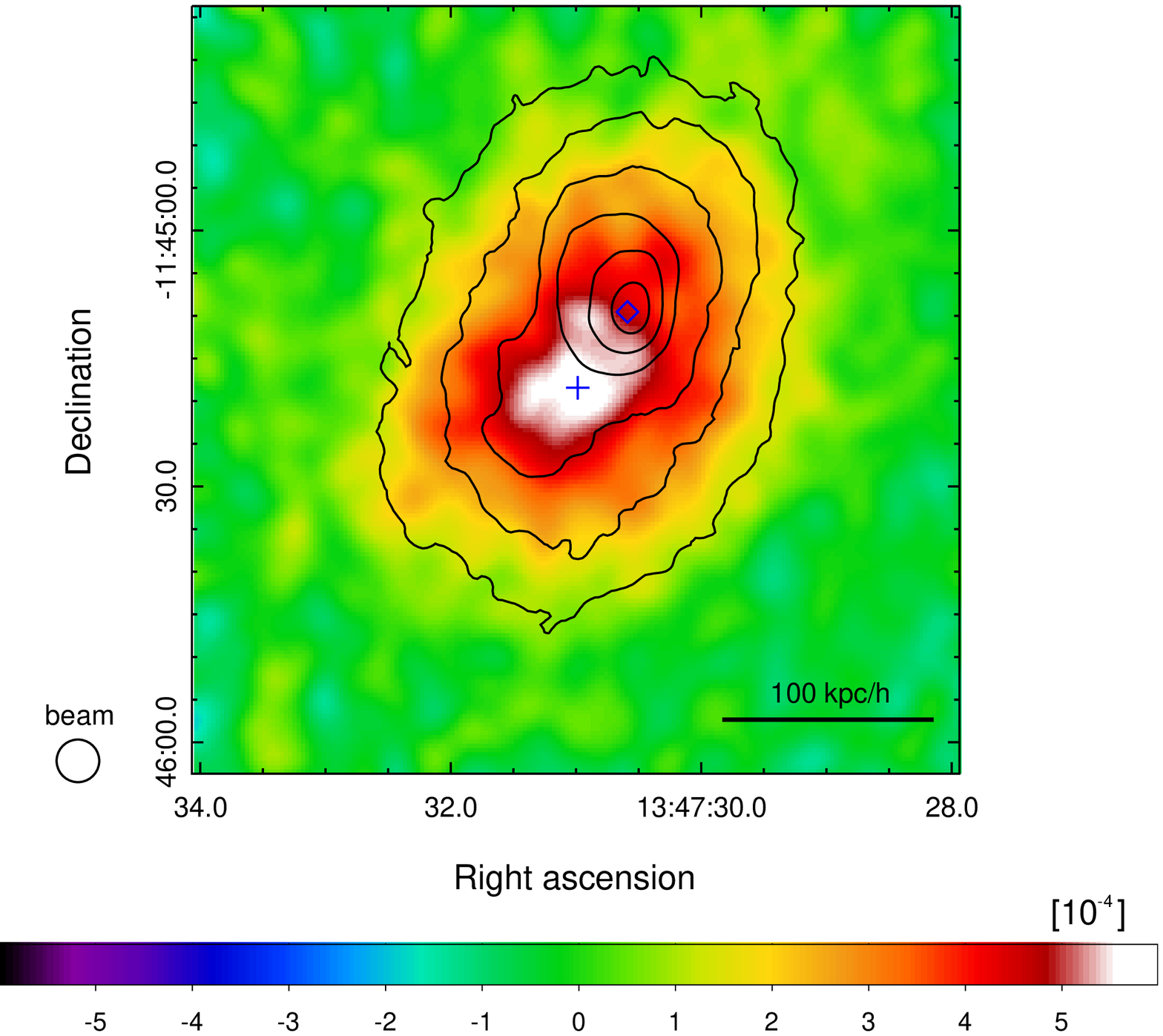} 
 \end{center}
\caption{Differential Compton $y$-parameter map of RX J1347.5--1145
  with $5''$ resolution reconstructed from the observed ALMA SZE map
  at 92~GHz. The map is corrected for the missing flux and the zero
  level is taken at $40''$ from the SZE intensity peak position marked
  by a cross. Overlaid for reference are the contours of the X-ray
  surface brightness at $0.4-7$ keV by Chandra corresponding to 64,
  32, 18, 8, 4, and 2$\%$ of the peak value, after being smoothed by a
  Gaussian kernel with $2.3''$ FWHM. The position of the subtracted
  AGN is marked by a diamond. }
\label{fig-yparam}
\end{figure}

We use the results of simulations presented in Section \ref{sec-sim}
to bridge between the observed ALMA map and the intrinsic signal of RX
J1347.5--1145. Figure \ref{fig-yparam} shows the Compton $y$-parameter
map reconstructed from the observed ALMA map in Figure \ref{fig-sz}.
The plotted values correspond to {\it differential} $y$-parameter
($\Delta y$) with respect to the positions at $40''$ from the SZE
intensity peak.  The relativistic correction by \citet{Nozawa05} has
been applied assuming the projected temperature shown in Figure
\ref{fig-xray}(b) at each sky position. The missing flux has been by
modeled by correction (ii) described in Section \ref{sec-sim}; the map
is divided by the mean reduction factor of $0.88$ ($c_1$ in equation
~[\ref{eq-conv}]) and the mean value at $40''$ from the emission peak
$\Delta y = (8.6 \pm 3.5) \times 10^{-6}$ is subtracted from the
entire map to define the zero level.

The peak value of the Compton $y$-parameter is $\Delta y = (5.9 \pm
0.4 \pm 0.8) \times 10^{-4}$ relative to the positions at $40''$ away
from the peak. Similarly, an integrated flux measured within the
radius $40''$ from the peak corresponds to $\Delta Y = \int \Delta y
d\Omega = (1.6 \pm 0.03 \pm 0.8)\times 10^{-11}$. In both cases,
quoted systematic errors are from the missing flux correction
(1.3~$\mu$Jy/arcsec$^2$) and the flux calibration (6\%).  If present,
the peculiar motion of the cluster along the line-of-sight changes the
above values further by $\sim 8\% ~[v_{\rm pec}/(1000 
\mbox{~km s$^{-1}$})][kT_{\rm e}/13 ~{\rm keV}]^{-1}$ via the kinematic SZE.

\subsection{Comparison with Previous SZE Observations}
\label{sec-compsz}

\begin{figure}
 \begin{center}
  \includegraphics[height=8cm]{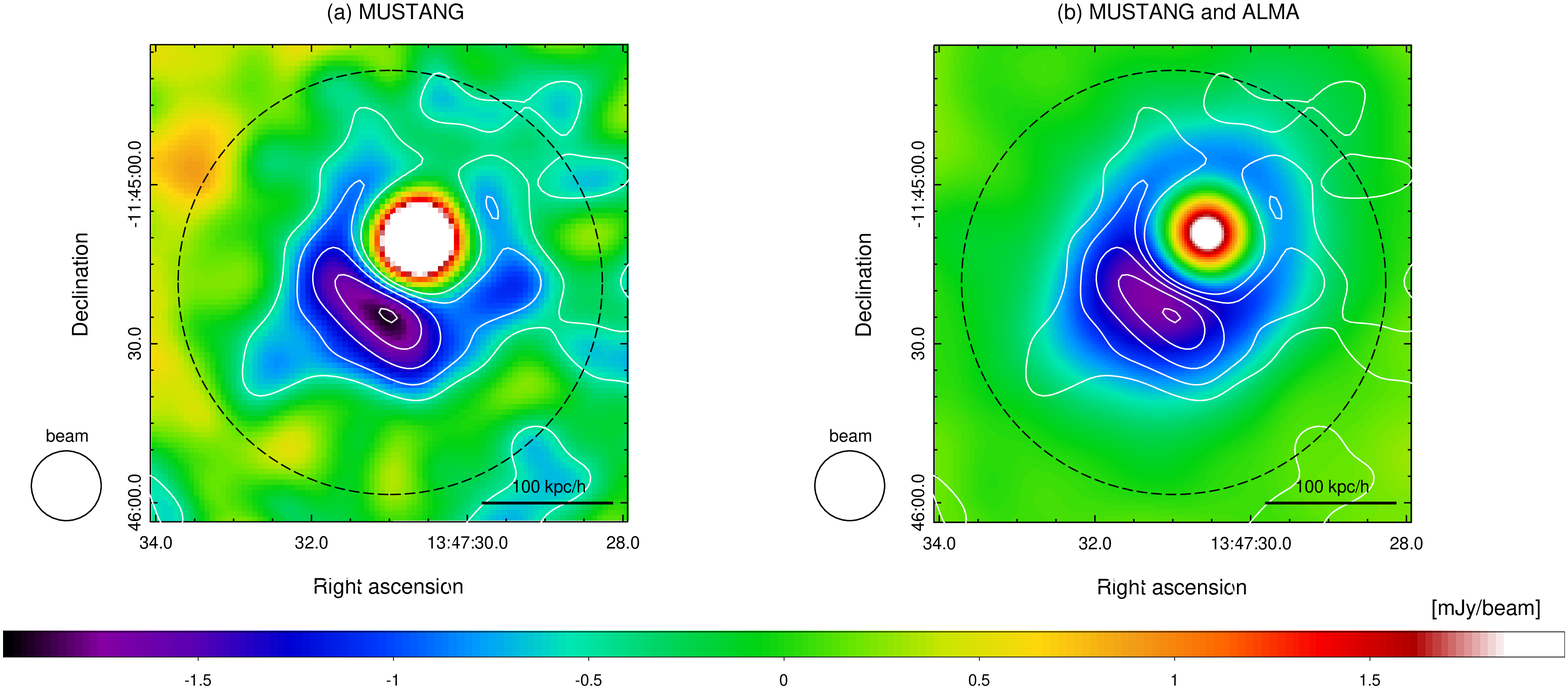}
  \includegraphics[height=8cm]{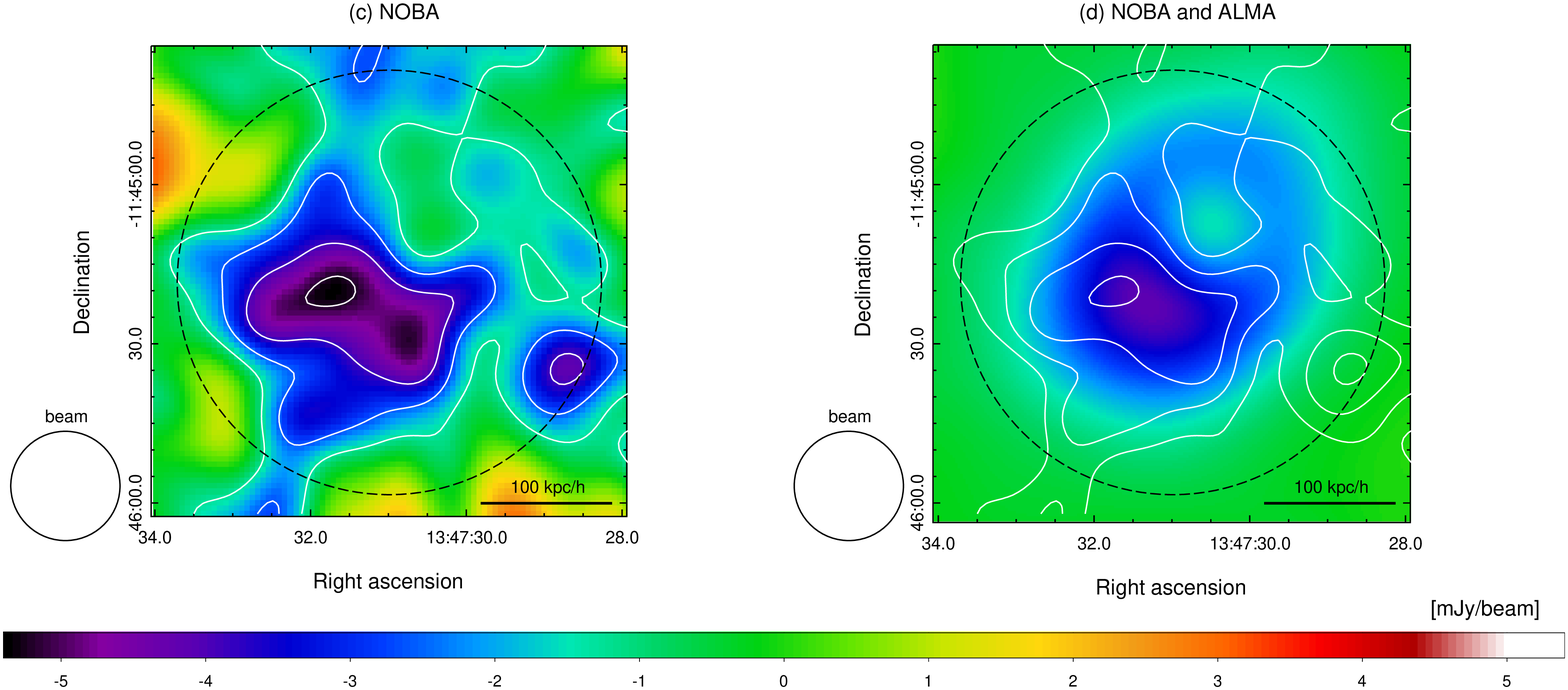} 
 \end{center}
\caption{Comparison of the deconvolved ALMA map with previous SZE
  observations of RX J1347.5--1145. (a) The MUSTANG map at 90~GHz
  \citep{Mason10,Korngut11} processed with an updated pipeline and
  smoothed to an effective beam size of $13.5''$ FWHM (B. Mason and
  C. Romero, private communication; \cite{Romero15}).  Contours
  indicate the S/N of $1-5$ in $1\sigma$ increments with $1\sigma \sim
  0.4$ mJy/beam depending on the exposure at each position of the
  map. (b) The overlay of the contours plotted in panel (a) on the
  ALMA map corrected for the missing flux, re-added the point source,
  and smoothed to $13.5''$ FWHM. (c) The NOBA map at 150~GHz
  \citep{Komatsu01} smoothed to an effective beam size of $20.6''$
  FWHM. Contours indicate the S/N of $1-4$ in $1\sigma$ increments
  with $1\sigma=1.3$ mJy/beam.  (d) The overlay of the contours
  plotted in panel (c) on the ALMA map converted to 150 GHz, corrected
  for the missing flux, re-added the point source, and smoothed to
  $20.6''$ FWHM. Dashed circles indicate the radius of $40''$ from the
  intensity peak in the unsmoothed ALMA map, at which the mean
  brightness values ($-0.046$~mJy/beam and $-0.76$~mJy/beam) in panels
  (b) and (d), are set equal to those of panels (a) and (c),
  respectively.}
\label{fig-compsz}
\end{figure}

Figure \ref{fig-compsz} exhibits comparison of the ALMA map with
previous SZE observations of RX J1347.5--1145 by MUSTANG
\citep{Mason10,Korngut11} and NOBA \citep{Komatsu01}. The MUSTANG data
shown in this figure are essentially the same as those published in
\citet{Mason10} and \citet{Korngut11}, except that they were processed
with an updated pipeline (B. Mason and C. Romero, private
communication; \cite{Romero15}). For closer comparison with the
MUSTANG data, the deconvolved ALMA map was corrected for the missing
flux by dividing by a factor of $0.88$ and adding a constant offset;
the central AGN with the inferred mean flux of $4.06$ mJy at 92 GHz
was re-added; the map was smoothed to the same effective resolution as
the MUSTANG contours; and the value of the offset is adjusted so that
the mean brightness at $40''$ from the SZE peak of the unsmoothed ALMA
map (the positions marked by the dashed circles in Figure
\ref{fig-compsz}) matches that of the MUSTANG map,
$-0.046$~mJy/beam. Note that we only compare the brightness relative
to these offset positions.

The NOBA data shown in Figure \ref{fig-compsz}(c) are identical to
those of \citet{Komatsu01}.  Also for direct comparison with the NOBA
data, the similar procedure to that mentioned above was applied to the
deconvolved ALMA map except that the ALMA map at 92 GHz was converted
to 150 GHz by multiplying a factor of 1.14 to correct for the SZE
spectrum (this factor includes the relativistic correction and is
accurate to better than $0.5\%$ at $kT_{\rm e}<25$ keV), the central
AGN was added with the flux expected at 150~GHz assuming a power-law
spectrum with the index $-0.527$ \citep{Sayers16}, and the mean
brightness at $40''$ from the peak is taken to be $-0.76$~mJy/beam.

Morphology of the ALMA map is in good agreement with both MUSTANG and
NOBA maps particularly in the south-east region, where the S/N of each
measurement is high.  It is also evident that eliminating the point
source contamination is crucial for recovering detailed structures of
the SZE.  The position of the south-east peak in the source-added and
smoothed maps (panels b and d) is offset by $\sim 7''$ compared to
that identified in the unsmoothed ALMA map (Figure \ref{fig-sz}). The
eastern ridge (the feature labeled as 2 in Figure 5 of \cite{Mason10})
is weaker than reported by \citet{Mason10} in both ALMA and updated
MUSTANG maps.  In accord with \citet{Mason10}, a low significance
decrement peak in the south-west quadrant of this cluster seen in the
NOBA map (the feature labeled as 3 in Figure 5 of \cite{Mason10}) is
absent in the ALMA map as well.

With respect to the positions marked by the dashed circles in Figure
\ref{fig-compsz}, the peak intensities of the decrement in the
source-added and smoothed ALMA maps shown in Figures
\ref{fig-compsz}(b)(d) are $-1.70 \pm 0.07 \pm 0.29$ mJy/beam and
$-3.39 \pm 0.13 \pm 0.66$ mJy/beam, respectively (the latter is a
converted value at 150~GHz); quoted systematic errors are from the
missing flux correction and the flux calibration. These intensities
are consistent with the corresponding values measured on the MUSTANG
map of $-1.9$ mJy/beam (with the $1\sigma$ statistical error of $\sim
0.4$ mJy/beam) and on the NOBA map of $-4.6 \pm 1.3 $ mJy/beam,
respectively, within the range of uncertainties in each measurement.

\section{Conclusions}
\label{sec-conc}

In this paper, we have presented the first image of the thermal SZE
obtained by ALMA.  The resulting angular resolution of $5''$
corresponds to $20 h^{-1}$kpc for our target galaxy cluster
RX~J1347.5--1145 at $z=0.451$. The present dataset achieves the
highest angular and physical spatial resolutions to date for imaging
the SZE.  The ALMA image has clearly resolved the bright central AGN,
the cool core, and the offsetted SZE peak in this cluster. It is in
good agreement with an electron pressure map reconstructed
independently from the X-ray data as well as with the previous SZE
observations of this cluster by NOBA \citep{Komatsu01,Kitayama04} and
MUSTANG \citep{Mason10,Korngut11}.
 
The statistical significance of the measurement has also improved
significantly; the achieved 1$\sigma$ sensitivity of the image is
0.017 mJy/beam or 0.12 mK$_{\rm CMB}$ at $5''$ FWHM.  The accuracy of
the map is limited primarily by missing flux arising from the lack of
short-spacing data in the current configuration of ALMA. We have
presented detailed analysis procedures including corrections for the
missing flux based on realistic imaging simulations for
RX~J1347.5--1145. We have shown that the structures up to the spatial
scale of $40''$ are faithfully recovered in the ALMA map.

Our results demonstrate that ALMA is a powerful instrument for imaging
the SZE in compact galaxy clusters with unprecedented angular
resolution and sensitivity.  They will also serve as guiding methods
for analyzing and interpreting future SZE images by ALMA.  
Completion of the Total Power Array for continuum observations as
  well as Band 1 receivers will significantly strengthen the
  capability of ALMA for imaging the SZE. Further implications of the
present results on the physics of galaxy clusters will be explored
separately in our forthcoming papers.

\begin{ack}
 
We thank Brian Mason and Charles Romero for providing the MUSTANG map
and helpful comments on the manuscript; Akiko Kawamura, Hiroshi Nagai,
and Kazuya Saigo for their support on the ALMA data reduction.  This
paper makes use of the following ALMA data:
ADS/JAO.ALMA\#2013.1.00246.S.  The scientific results of this paper
are based in part on data obtained from the Chandra Data Archive:
ObsID 506, 507, 3592, 13516, 13999, and 14407.  ALMA is a partnership
of ESO (representing its member states), NSF (USA) and NINS (Japan),
together with NRC (Canada) and NSC and ASIAA (Taiwan), in cooperation
with the Republic of Chile. The Joint ALMA Observatory is operated by
ESO, AUI/NRAO and NAOJ. The National Radio Astronomy Observatory is a
facility of the National Science Foundation operated under cooperative
agreement by Associated Universities, Inc. This work was supported by
the Grants-in-Aid for Scientific Research by the Japan Society for the
Promotion of Science with grant numbers 24340035 (Y.S.), 25400236
(T.K.), 26400218 (M.T.), 15H02073 (R.K.), 15H03639 (T.A.), 15K17610
(S.U.), and 15K17614 (T.A.).  T.K. was supported by the ALMA Japan
Research Grant of NAOJ Chile Observatory, NAOJ-ALMA-0150.

\end{ack}

\end{document}